\documentclass[preprint2,times,tighten]{aastex6}
\pdfoutput=1
\usepackage{amsmath,amstext}
\usepackage[all]{hypcap} 

\newcommand{\hmsun}{{h^{-1}}{\rm M}_{\solar}}
\newcommand{\solar}{_{\mathord\odot}}
\newcommand{\hmpc}{\ifmmode{h^{-1}{\rm Mpc}}\;\else${h^{-1}}${\rm Mpc}\fi}
\newcommand{\hpc}{\ifmmode{h^{-1}{\rm pc}}\;\else${h^{-1}}${\rm pc}\fi}
\newcommand{\hMpc}{\ifmmode{h^{-1}{\rm Mpc}}\;\else${h^{-1}}${\rm Mpc}\fi}
\newcommand{\hGpc}{\ifmmode{h^{-1}{\rm Gpc}}\;\else${h^{-1}}${\rm Gpc}\fi}
\newcommand{\hkpc}{\ifmmode{h^{-1}{\rm kpc}}\;\else${h^{-1}}${\rm kpc}\fi}

\newcommand{\msun}{{\rm M}_{\solar}}

\newcommand{\mr}{\ifmmode{M_r}\;\else$M_r$\fi}
\newcommand{\rd}{\ifmmode{R_\delta}\;\else$R_\delta$\fi}
\newcommand{\ngals}{\ifmmode{N_{\rm gals}}\;\else$N_{\rm gals}$\fi}

\newcommand{\kms}{\,\rm{km}\,\rm{s}^{-1}}

\usepackage{soul}

\shorttitle{Host halo and satellites of the Milky Way}
\shortauthors{Lu et al.}

\begin{document}

\title{The connection between the host halo and the satellite galaxies of the Milky Way} 
\author{Yu~Lu\altaffilmark{1}}
\author{Andrew~Benson\altaffilmark{1}}
\author{Yao-Yuan~Mao\altaffilmark{2}}
\author{Stephanie~Tonnesen\altaffilmark{1}}
\author{Annika~H.~G.~Peter\altaffilmark{3}}
\author{Andrew R. Wetzel\altaffilmark{1,4}}
\author{Michael~Boylan-Kolchin\altaffilmark{5}}
\author{Risa H. Wechsler\altaffilmark{2}}
\altaffiltext{1}{The Observatories, The Carnegie Institution for Science, 813 Santa Barbara Street, Pasadena, CA 91101, USA}
\altaffiltext{2}{Kavli Institute for Particle Astrophysics and Cosmology; Department of Physics, Stanford University, Stanford, CA 94305, USA;\\ SLAC National Accelerator Laboratory, Menlo Park, CA, 94025, USA}
\altaffiltext{3}{CCAPP and Department of Physics, The Ohio State University, 191 W.\ Woodruff Ave., Columbus, OH 43210, USA;\\ Department of Astronomy, The Ohio State University, 140 W.\ 18th Ave., Columbus, OH 43210, USA}
\altaffiltext{4}{TAPIR, California Institute of Technology, Pasadena, CA 91125, USA}
\altaffiltext{5}{Department of Astronomy, The University of Texas at Austin, 2515 Speedway, Stop C1400, Austin, TX 78712-1205, USA}

\begin{abstract}

Many properties of the Milky Way's dark matter halo, including its mass assembly history, concentration, and subhalo population, remain poorly constrained.
We explore the connection between these properties of the Milky Way and its satellite galaxy population, especially the implication of the presence of the Magellanic Clouds for the properties of the Milky Way halo. 
Using a suite of high-resolution $N$-body simulations of Milky Way-mass halos with a fixed final $M_{\rm vir} \sim 10^{12.1}\msun$, we find that the presence of Magellanic Cloud-like satellites strongly correlates with the assembly history, concentration, and subhalo population of the host halo, such that Milky Way-mass systems with Magellanic Clouds have lower concentration, more rapid recent accretion, and more massive subhalos than typical halos of the same mass.
Using a flexible semi-analytic galaxy formation model that is tuned to reproduce the stellar mass function of the classical dwarf galaxies of the Milky Way with Markov-Chain Monte-Carlo, we show that adopting host halos with different mass-assembly histories and concentrations can lead to different best-fit models for galaxy-formation physics, especially for the strength of feedback. These biases arise because the presence of the Magellanic Clouds boosts the overall population of high-mass subhalos, thus requiring a different stellar-mass-to-halo-mass ratio to match the data. 
These biases also lead to significant differences in the mass--metallicity relation, the kinematics of low-mass satellites, the number counts of small satellites associated with the Magellanic Clouds, and the stellar mass of Milky Way itself.  Observations of these galaxy properties can thus provide useful constraints on the properties of the Milky Way halo.

\end{abstract}


\section{Introduction}\label{sec:introduction}

The Milky Way (MW) galaxy and its satellite galaxies provide an excellent 
laboratory for constraining cosmology and galaxy formation physics.
Accurately modeling the formation of galaxies in the MW system requires stringent 
constraints on galaxy formation physics and the properties of the dark matter halos in this particular environment. 
One of the unusual characteristics of the MW galaxy is that it has two massive satellite 
galaxies, the Large and Small Magellanic Clouds (MCs). 
Both of them are measured to have maximum circular velocities $V_{\rm max} \geq 55 {\rm km~s}^{-1}$ with 
magnitudes $M_{\rm v}=-18.5$ and $-17.1$, respectively \citep[e.g.,][]{van-den-Bergh2000a, van-der-Marel2002a, Stanimirovic2004a, van-der-Marel2014a}. 
The next brightest satellite is Sagittarius dSph, about 4 magnitudes dimmer, 
with $V_{\rm max} \sim 20 {\rm km~s}^{-1}$ \citep{Strigari2007a}. 
In addition, the MCs have significantly higher stellar masses than the rest of the MW 
satellite galaxy population. 
The stellar mass of the LMC is about $1.5\times10^{9}\msun$,  
and the stellar mass of the SMC is about $4.6\times10^{8}\msun$, which are 1.8 and 1.3 dex higher than 
the stellar mass of the core of Sagittarius, which is $2.1\times10^7\msun$ \citep{McConnachie2012a}.
Several works have shown that satellites as bright as the LMC are rare around MW-mass galaxies \citep[e.g.,][]{Liu2011a, Guo2011a, Tollerud2011a}.
This rarity suggests that the presence of such high-mass satellites may have important implications for the formation history of the MW halo.
In this paper, we will study the connections between the properties of the MW 
host halo and the existence of MCs, and explore the impact of the host halo properties on 
the inference of galaxy formation using observations of the satellite galaxies of the MW, including the MCs.

There have been a number of theoretical studies in the literature investigating the implications of 
high-mass substructures in the MW analogues. 
\citet{Boylan-Kolchin2011b} used a large sample of MW-mass halos
($4.3\times10^{11} \leq M_{\rm vir}/ \msun \leq 4.3\times 10^{12}$)
extracted from the 
Millennium-II simulation and found that 20\% of MW-mass halos host an LMC or SMC, 
and that only $\sim2.5$\% of such halos host both MCs as a binary. This indicates that the MW system must be 
rare in a Cold Dark Matter (CDM) Universe. 
A similar conclusion was reached by \citet{Busha2011a} using a different set of simulations. 

On the other hand, selecting MW analogues from cosmological simulations by requiring the existence 
of MCs can be used to constrain the MW's virial mass, the dark matter halo density profile, and 
its satellite accretion history (e.g., \citealt{Busha2011b, Cautun2014b}). 
In recent studies, \citet{Wang2015a} found that halo concentration strongly affects the estimate of halo mass 
using dynamical tracers of the MW. 
\citet{Mao2015a} demonstrated that halo concentration influences the inferences for the mass 
and other properties of the MW halo from satellite occupation statistics. 
It has also been shown that the abundance of subhalos correlates with the mass-assembly 
history of the halo \citep[e.g.,][]{Zentner2005a, Zhu2006a, Mao2015a}.

The baryonic processes of galaxy formation also significantly impact the properties of  the satellite galaxy population \citep{Bullock2001b, Benson2002a, Benson2002b, Font2011a, Gomez2014a, Sawala2014a, Wetzel2016a}. A number of authors have adopted various approaches to model the formation of MW satellite galaxies.  It has been shown that satellites as bright as the LMC are rarely predicted for MW-mass galaxies, with only $\sim10$\% 
of the MW-mass galaxies having satellites as bright as the LMC \citep[e.g.,][]{Kauffmann1993a,  Koposov2009a, Liu2011a}.
Moreover, \citet{Koposov2009a} found that an unusually high star formation efficiency was needed in their model to reproduce objects as bright as LMC and SMC, implying again that the MW system is unusual owing to the existence of MCs. Similarly, \citet{Okamoto2010a} explored a range of feedback models to add galaxies to some of the high-resolution Aquarius halos \citep{Springel2008a}, and again found it difficult to readily reproduce galaxies with luminosities as high as the MC's while simultaneously matching the luminosities of lower mass satellite galaxies. The same conclusion was reached by \citet{Starkenburg2013a} using a different model. 

While this difficulty could stem from incomplete physics being considered in current galaxy formation models, it is possible that the influence of the underlying dark matter structure is the cause of the issue. As the backbone of galaxy formation, the dark matter halo and its associated subhalos have a strong impact on the properties of the hosted galaxies. For example, \citet{Starkenburg2013a} (and references therein) showed a strong correlation between the number of satellites and the dark matter mass of the host halo. 
It has also been shown that the formation of the satellite galaxies of the MW is highly stochastic and sensitively depends on the subhalo population \citep[e.g.,][]{Cautun2014b, Guo2015a}.
Also, \citet{Gomez2012a} found that halo merger histories and galaxy formation physics are degenerate under certain observational data constraints.
Thus, to accurately model the formation of a particular galaxy like the MW, it is important to understand how the prior information about the properties of 
the dark matter halo, including its concentration, mass-assembly history, and hosted subhalos, 
affects the model predictions of the MW and its satellite populations. 
Such an understanding is also important for interpreting observational data on the Andromeda galaxy and  MW analogues in the more distant Universe 
\citep[e.g.,][]{Liu2011a, Guo2011a, Tollerud2011a, Nierenberg2013a, Nierenberg2016a}. 

This paper is dedicated to investigating the host halo prior. 
First, we attempt to gain insight into what halo properties are of interest when modeling the formation of MW satellite galaxies. 
To achieve this, we study a set of $N$-body simulations of MW-mass halos, and apply a semi-analytic model to the merger trees extracted from those simulations to study galaxy properties. 
To understand how the halo prior influences inferences of galaxy formation physics, we employ a Markov-Chain Monte-Carlo (MCMC) machinery that is joint with our SAM to explore the parameter space of a galaxy formation model. 
Second, using the constrained models, we investigate the effect of the halo prior and gain insight into which aspects of modeling and observational work are needed to further tighten the constraints on the formation of the MW and its satellite galaxies. 
Specifically, we explore the parameter space in the galaxy formation model to understand the constraining power of observational estimates  of the stellar mass-metallicity relation, the kinematics of satellite galaxies, and the number counts of small satellite galaxies brought in by the MCs. 
In addition, extrapolating the constrained model to central galaxies of MW-mass halos, we demonstrate how the connections between the host halo and the satellite galaxies can also influence the mass of the central galaxies, shedding light on understanding the effects of the halo assembly bias \citep[e.g.][]{Gao2005a, Wechsler2006a, Jing2007a} for galaxy formation.

In this paper, we describe the simulations and the SAM adopted in \S\ref{sec:method}. 
The results on the relation between the host halo properties and the high-mass subhalos 
from analyzing the dark matter simulations are presented in \S\ref{sec:mah}. 
We present the model inferences based on the SAM from the MW satellite stellar mass function 
in \S\ref{sec:inference}.  
We summarize the conclusions of the study and discuss the implications in \S\ref{sec:discussion}.

\section{Methodology}\label{sec:method}

\subsection{The simulations}\label{sec:simulation}
In this study, we adopt two sets of $N$-body simulations, a suite of high-resolution 
zoom-in simulations of MW-mass halos \citep{Mao2015a} as our primary halo sample, 
and a cosmological simulation \texttt{c125-2048} to increase 
the sample size for better statistics. 
The \texttt{c125-2048} simulation is a dark matter-only cosmological simulation run with 
$2048^3$ particles and a side length of $125 ~\hMpc$, particle mass of 
$1.8\times10^7 ~\hmsun$, started at $z=199$.  
The softening length is $0.5 ~\hkpc$, constant in comoving scale. 
The cosmological parameters are $\Omega_{\rm M} = 0.286$, $\Omega_{\Lambda} = 0.714$, 
$h = 0.7$, $\sigma_8 = 0.82$, and $n_s = 0.96$. This simulation was used previously 
in \cite{Mao2015a} and \cite{Desmond2015a}.
The zoom-in simulations consist of 46 halos selected for resimulation from a parent 
simulation \texttt{c125-1024}, a lower resolution sister simulation of \texttt{c125-2048}. 
The parameters and initial conditions of these two boxes are identical, but 
\texttt{c125-1024} contains $1024^3$ particles and starts at a different redshift, $z=99$. 
All the zoom-in simulation halos fall in the final halo mass range 
$M_{\rm vir}=10^{12.1\pm0.03}\msun$, where the virial mass definition follows 
\citet{Bryan1998a}. 
The mass of the particles in the zoom-in simulations is $3.0 \times 10^5\,\hmsun$. 
The softening length in the highest-resolution region is $170 \, \hpc$ comoving. 
Out of the 46 zoom-in simulation halos, we adopt 38 of them in this paper and discard the other 8 halos because the latter were selected purposely from halos with very large Lagrangian volumes, which render the MCMC-SAM 
calculation too computationally expensive. 
When larger samples are needed to enhance statistical significance, we adopt
the \texttt{c125-2048} cosmological simulation and select all halos in the same virial mass 
range $M_{\rm vir}=10^{12.1\pm0.03}\msun$ at $z=0$, which yields a large sample of 
$\sim1300$ halos.
This halo mass is consistent with many observational constraints of the halo mass 
of the MW \citep{Cautun2014b, Eadie2015a, Xue2008a, Gonzalez2013a}.
We restrict our study to this halo mass range as chosen in \citep{Mao2015a} 
to limit the variation in the subhalo populations introduced by varying halo mass, since varying halo mass will change the statistical properties of 
hosted subhalos \citep[e.g.][]{Kravtsov2004b, Wang2012a, Cautun2014b}. 

For each of these simulations, we use the halo catalog generated by the \textsc{Rockstar} halo finder \citep{Behroozi2013a} and merger trees generated by the \textsc{Consistent Trees} merger tree code \citep{Behroozi2013b}.
We adopt the virial overdensity (${\Delta}_{\rm vir}$) as our halo mass definition \citep{Bryan1998a}.   \citet{Mao2015a} tested the numerical convergence of the subhalo circular velocity function 
and found that a conservative lower limit for convergence on the maximum halo velocity is $50 \kms{}$ for 
the \texttt{c125-2048} cosmological simulation and $10 \kms{}$ for the zoom-in MW simulations. 
The halo concentration parameter used in this study is calculated by fitting the Navarro--Frenk--White (NFW) profile to the dark matter density distribution \citep{Navarro1997a}.
For further details of these simulations, including halo identification and merger tree 
construction, readers are referred to \citet{Mao2015a}.

\subsection{The semi-analytic model}\label{sec:sam}

To study the baryonic component of MW satellites, we adopt a SAM 
developed by \citet{Lu2011b}, in which the parameterizations for the baryonic processes 
of galaxy formation are generalized to encompass a wide range of possibilities. 
In this study, we apply this model to the merger trees extracted from the zoom-in simulations. 
Owing to the high mass resolution of the simulations, the merger trees include a 
large number of low-mass progenitor halos that form satellite galaxies below 
the mass range probed in this study. 
Following the full merger tree in a SAM calculation would be inefficient for this study.
To allow exhaustive parameter space exploration, we need to adopt an approximation
to reduce the computation time of the model without significant loss of accuracy. 
In this paper, we adopt a scheme where we focus the computation only on the subhalos 
that are still present at $z=0$ to be able to compare the predictions with observational data. 
We safely ignore the progenitor halos that have been tidally disrupted by $z=0$. 
These progenitor halos are accreted into the final host very early and may have donated 
their stellar mass to the central galaxy or the diffuse stellar halo. 
Because these components are not studied in this paper, we simply ignore these processes 
in this paper. 
Moreover, because these ignored halos only form tiny amount of stars, their effects on the 
chemical abundances and radiative cooling for the descendant galaxies are also negligible.

Some of those disrupted subhalos may leave an ``orphan'' galaxy (satellite galaxy without 
a host subhalo above the resolution limit).
Checking with our fiducial model, we find that the predicted stellar masses of the ``orphan'' 
galaxies are significantly below the mass range probed in this paper, 
owing to the high mass resolution of the simulations. 
We note that subhalos that are accreted into subhalo branches are not ignored in this scheme,
even if they may have been disrupted by $z=0$, because they may contribute their stellar mass 
to the satellite galaxies in which we are interested.  
We have compared the stellar masses of the satellite galaxies predicted using this scheme 
and those predicted with the full SAM, which follows the entire tree.
We find that $<4$\% of the satellite galaxies with $M_*\geq10^4\msun$ experience a deviation in mass of more than 2\% from the scheme following the full merger trees. 
We conclude that the scheme accurately reproduces the stellar masses of all satellites 
in the mass range of this study. 
We therefore adopt the scheme in this study. 
This scheme typically reduces the computation time by a factor of 4 for each MW merger tree. 
The benefit in speeding up the calculation allows us to use MCMC 
to sample the posterior distribution of the model under data constraints of the MW 
satellite stellar mass function.


\section{Host halo and subhalos}\label{sec:mah}

\begin{figure*}[htb!]
\begin{center}
\includegraphics[width=0.95\textwidth]{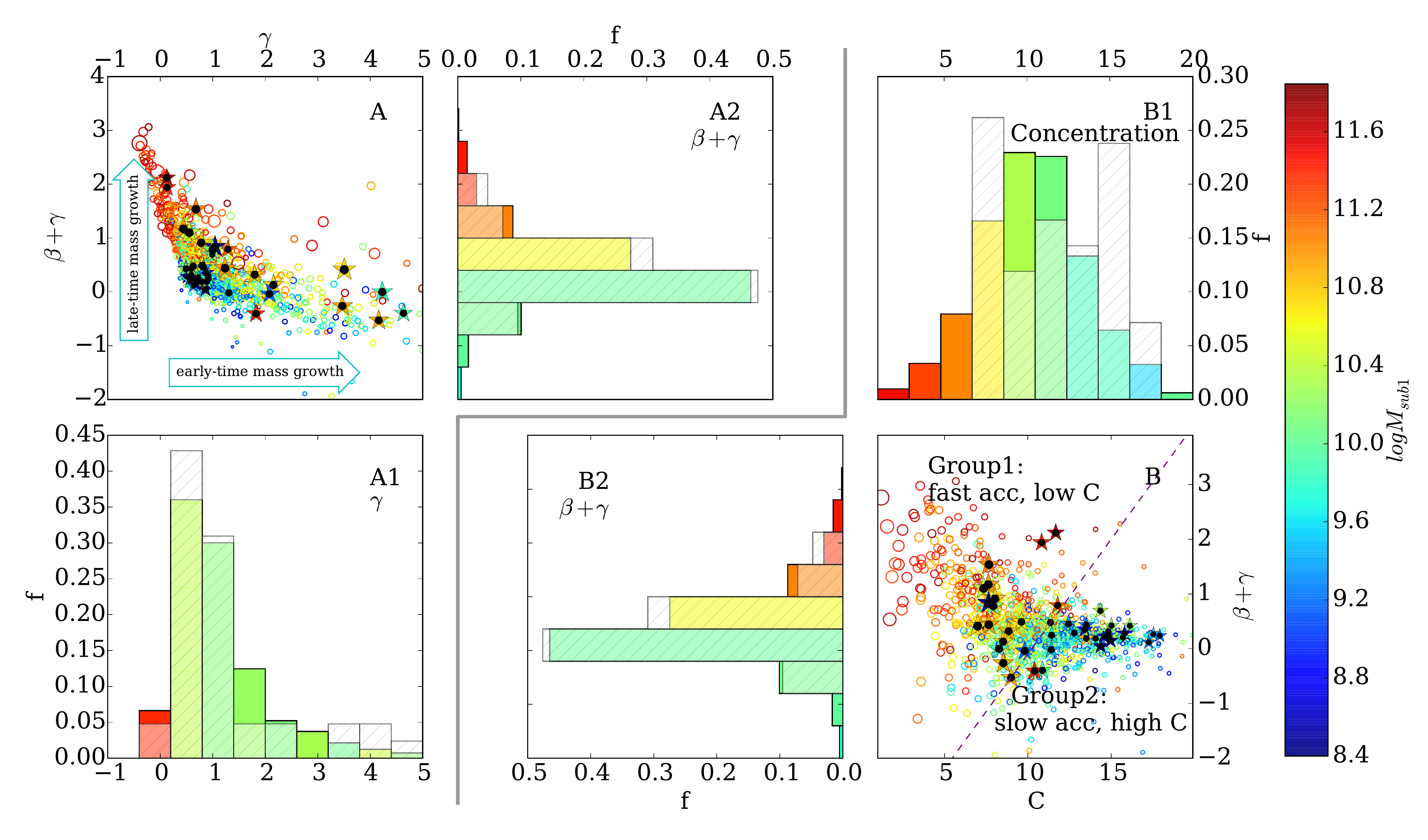}
\caption{The distribution of simulated MW hosts in parameter space. 
Panel {\bf A} shows the distribution of the host halo mass-assembly history 
fitting parameters, $\gamma$ and $\beta+\gamma$. 
The parameter $\gamma$ indicates the early accretion rate, and $\beta+\gamma$ 
indicates the late accretion rate. 
Each circle represents a MW host from the cosmological simulation \texttt{c125-2048} 
and each star with a central black circle represents a zoom-in simulation. 
The adjacent panels, {\bf A1} and {\bf A2}, show the marginalized distribution of the halos 
in $\gamma$ and $\beta+\gamma$, respectively.  
Panel {\bf B} shows the distribution in the space of the concentration parameter, $c$, 
versus $\beta+\gamma$. 
Similarly, Panels {\bf B1} and {\bf B2} show the marginal distribution of these parameters. 
The median concentration is $c_{\rm median} = 10.4$ for the cosmological samples, 
and $c_{\rm median}=11.5$ for the zoom-in simulation samples. 
The colors in Panels A and B represent the mass of the most massive 
subhalo hosted by each MW halo, as indicated in the color bar on the right.
The solid color histograms show the distribution of the cosmological simulation halos, 
while the hashed gray histograms represent the distribution of the zoom-in halos. 
The color of the histograms indicates the median mass of the most massive subhalos in each bin. 
The size of each symbol in Panels A and B is proportional to the
NFW core radius $r_{\rm s}$ (or inversely proportional to the halo concentration parameter, $c$). 
The purple dashed line in Panel B splits the distribution into the two groups used to
explore the halo prior in this paper.  
}
\label{fig:haloparams}
\end{center}
\end{figure*}

\begin{figure*}[htb!]
\begin{center}
\includegraphics[width=0.80\textwidth]{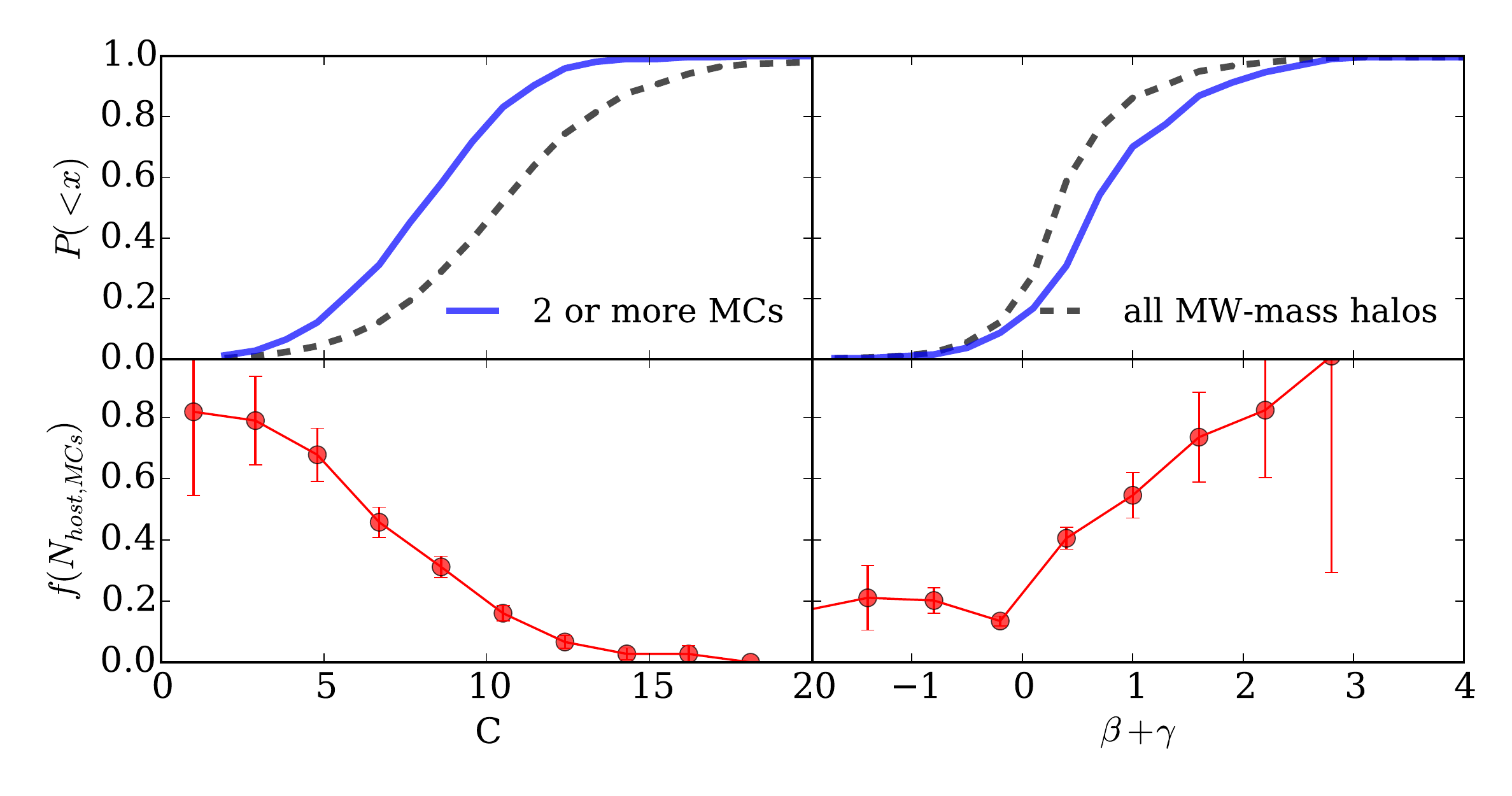}
\caption{{\bf Upper}: The cumulative distribution functions of the concentration parameter $c$ ({\bf upper-left}) 
and the MAH fitting parameter $\beta+\gamma$ ({\bf upper-right}) of the MW-mass halos 
in the cosmological simulation. The black dashed line in each upper panel denotes 
the distribution function for all the MW-sized halos, while the blue solid line denotes 
the distribution function for the halos that host 2 or more subhalos with $V_{\rm max}>55 \kms$. 
{\bf Lower}: The fraction of MW-mass halos that host at least two subhalos with 
$V_{\rm max}>55 \kms$ as a function of the concentration of the host ({\bf lower-left}) 
and the accretion history parameter $\beta+\gamma$ ({\bf lower-right}).
The error bars represent the standard deviation of a Poisson distribution.
MW-mass halos that host MC-mass satellites are more likely to have low concentration and more rapid recent mass accretion.
}
\label{fig:fnsub_vcut}
\end{center}
\end{figure*}

In this section, we study the correlation between the existence of a high-mass satellite galaxy like the LMC, and host halo properties. 
With the simulated dark matter halos, we first study their mass accretion histories (MAHs). 
The MAH is defined as the virial mass of the main-branch halo as a function of redshift or time, 
with the virial mass definition following \citet{Bryan1998a}.
To capture the overall shape of the MAHs, we adopt the two-parameter model proposed by \citet{Tasitsiomi2004a}. 
This model combines the exponential form of \cite{Wechsler2002a} 
and the power-law form of \citet{van-den-Bosch2002a}, 
\begin{equation}\label{eqn:mah_mcb}
M(z)=M_0 \left(\frac{1+z}{1+z_0}\right)^{-\beta} \exp\left[{-\gamma (z-z_0)}\right],
\end{equation}
where $\beta$ and $\gamma$ are model parameters determining the shape of the MAH, and
$M_0$ is the normalization for the halo mass at $z=z_0$. 
Using this model, \citet{McBride2009a} fit the MAHs of halos in the Millennium simulation 
in a large halo mass range, and found that this two-parameter model captures the halo MAHs 
remarkably well.\footnote{\citet{McBride2009a} use the opposite sign convention for $\beta$ compared to what we adopt here.} 
Following the previous work, we also fit the MAHs of the MW halos in our simulations using this model. 
A merit of this fitting function is that the model parameters characterize 
the accretion rate of a halo at early and late epochs. 
In this model, we find that 
\begin{equation}
\frac{{\rm d} \log M} {{\rm d}  z}=-\left(\frac{\beta}{1+z} +\gamma \right),
\end{equation}
and 
\begin{equation}
\frac{{\rm d} \log M}{{\rm d} \log a }= \beta + {\gamma (1+z)}.
\end{equation}
The derivations show that at the present time, when $z=0$ and $a=1$, the accretion 
rate ${\rm d} \log M/{\rm d} \log a$ is characterized by $\beta+\gamma$,
and at early times, the accretion rate is characterized by $\gamma$.
Motivated by these indications, we will use $\beta+\gamma$ and $\gamma$ extracted from the best-fit model 
to characterize the accretion rate at late and early times for each halo, and to study 
the correlations between the MAH and the satellite population.

In Panel A of Figure \ref{fig:haloparams}, we show the distribution of all the simulated 
MW host halos in the parameter space defined 
by the MAH fitting parameters, $\beta+\gamma$ and $\gamma$.  
The halos from the cosmological box (\texttt{c125-2048}) are shown as circles, 
and the halos from the zoom-in simulation are shown as stars with a black circle at the center. 
All the halos populate a particular region of the parameter space with a mode located  
around $\beta+\gamma=0.3$ and $\gamma=0.7$. 
This distribution is in agreement with what is found in other simulations \citep{McBride2009a, Taylor2011a}. 
In the diagram, halos in the upper left branch have rapid late-time accretion, 
while those in the lower-right branch have rapid early-time accretion.

The size of the symbols in Figure \ref{fig:haloparams} is linearly proportional to the scale radius, $r_{\rm s}$, of the NFW profile, and inversely proportional to the concentration parameter $c=R_{\rm vir}/r_{\rm s}$, 
where $R_{\rm vir}$ is the virial radius of the halo. 
From the figure, it can be seen that the concentrations are correlated with the MAH fitting parameters.  
In general, halos with faster late-time accretion tend to have lower concentration 
(or larger $r_{\rm s}$ for given $R_{\rm vir}$). 
This is more clearly seen in Panel B of Figure \ref{fig:haloparams}, where we plot the 
distribution of the simulated host halos in a diagram defined by halo concentration $c$ and $\beta+\gamma$.
In the diagram, one can find that rapid accreting halos also tend to have a lower concentration
and hence tend to be located in the upper-left part of the diagram.

In the figure, we also color code each halo according to the mass of its most massive subhalo, 
$M_{\rm sub1}$, which is defined by the total mass confined in the phase space of the subhalo 
with the \textsc{Rockstar} halo finder \citep{Behroozi2013b}.  
The scale of $\log M_{\rm sub1}$ and the corresponding color are shown in the color bar on the 
right side of Figure \ref{fig:haloparams}.
An interesting trend shown in the figure is that the mass of the most massive subhalo is also 
correlated with the MAH parameters and the concentration of the host. 
Hosts that have more rapid late-time accretion and lower concentration are more likely to host a high mass 
subhalo. 
We note that when we choose to use other subhalo properties 
as a proxy for high-mass satellites, for instance the peak subhalo mass ($M_{\rm peak}$), the subhalo mass at accretion ($M_{\rm acc}$), or the maximum subhalo circular velocity ($V_{\rm max}$), the trends we describe here still hold, 
but become weaker \citep[see also][]{Mao2015a}. This is because these quantities more closely reflect the accretion of a high-mass subhalo
at a much earlier time. 
The massive stellar streams or debris of tidally disrupted satellite galaxies may 
also contain information about the properties of the MW host halo, but the instantaneous mass of 
high-mass subhalos is more closely related to the present-day concentration and the recent accretion history of the host.

Adjacent to Panels A and B, we show the marginalized distribution of $\gamma$, 
$\beta+\gamma$, and $c$ in Panel A1, A2, and B1, respectively. 
The distribution of $\beta+\gamma$ is repeated in Panel B2.  
The solid color bars represent the distribution of the cosmological simulation halos. 
The colors of each bar indicates the median mass of the most massive subhalo for all the hosts in the bin.
Again, we can see a clear sequence that halos with higher $\beta+\gamma$ and a lower concentration 
tend to host a higher mass subhalo.
The gray hashed bars represent the zoom-in simulation halos. 
The small sample size of the zoom-in simulation covers the distribution of 
the population reasonably well with some discrepancy in the halo concentration distribution. 

\begin{figure*}[htb!]
\begin{center}
\includegraphics[width=0.95\textwidth]{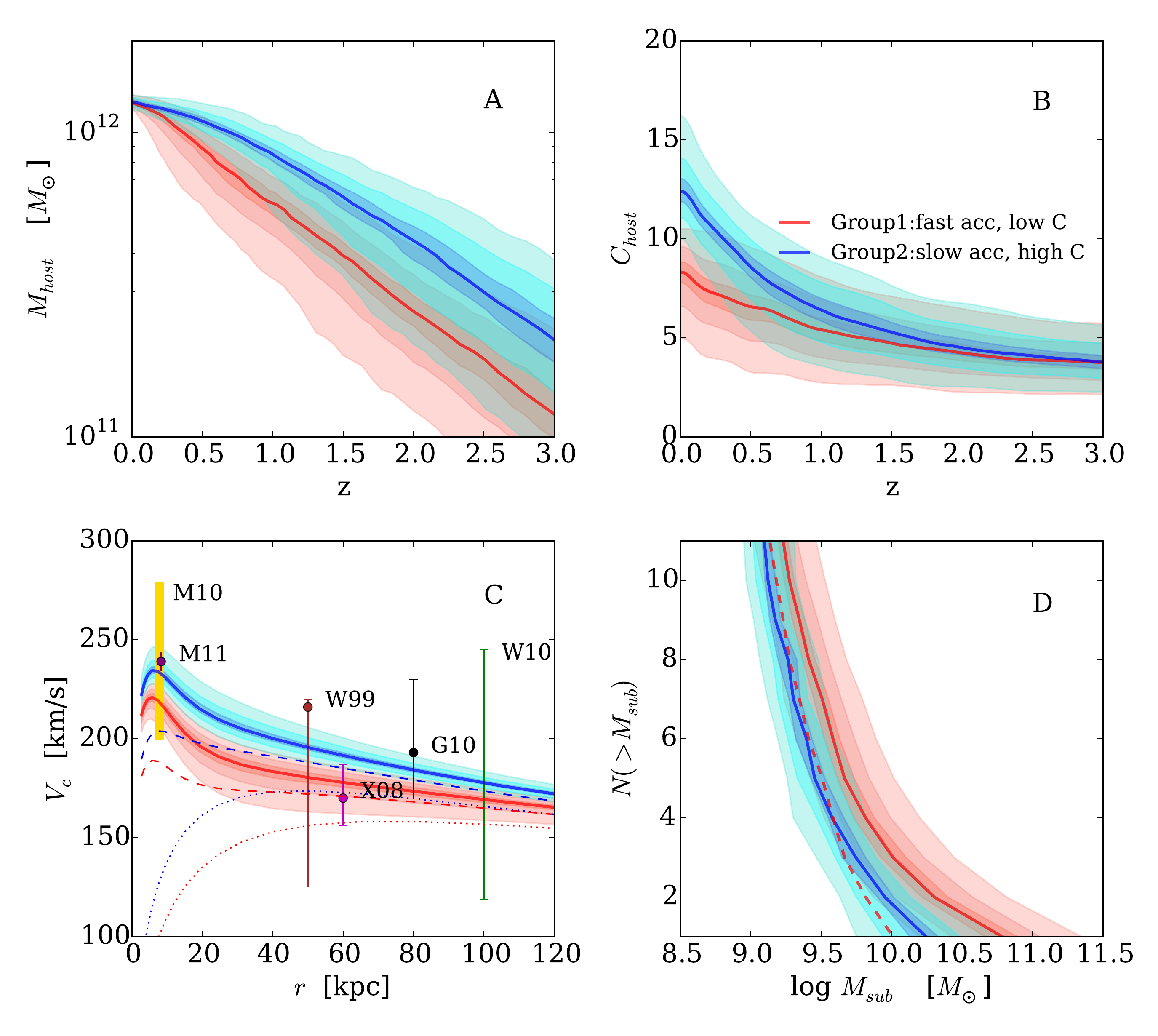}
\caption{Overall comparisons between the two host halo groups. 
The host mass-assembly history (Panel {\bf A}), the evolution of concentration (Panel {\bf B}), 
the rotation curve (Panel {\bf C}),
and the subhalo mass function (Panel {\bf D}) of MW-mass hosts predicted by the cosmological simulation. 
The red is for host halos in Group 1, which have rapid recent accretion and low concentration, and the blue is for host halos in Group 2, which have slow recent accretion and high concentration.  
The bands with decreasing intensity encompass the 20\%, 50\%, and 80\% of the distribution for each halo group. 
The solid lines in the middle of the bands denote the medians.  
In Panel {\bf C}, the predicted rotation curves shown by the color bands taking into account 
the dark matter halo, stellar disk and bulge, and the effect of adiabatic contraction. 
The dotted lines denote the predictions 
for dark matter only using the NFW formula with the median 
concentrations, $c=8.2$ for Group 1 hosts and $c=12.5$ for Group 2 hosts. 
The dashed lines denote the predictions for the NFW dark matter halo profile plus 
stellar disk and bulge, but the halo contraction is ignored. 
The data points and error bars show various observational constraints of 
\citet{McMillan2010a}--M10, \citet{McMillan2011a}--M11, \citet{Wilkinson1999a}--WE99, \citet{Xue2008a}--X08,
\citet{Gnedin2010a}--G10, and \citet{Watkins2010a}--W10.
The red dashed line in Panel {\bf D} denotes the median of the subhalo mass function of Group 1 halos 
shifted down by 2. 
}
\label{fig:halo_groups}
\end{center}
\end{figure*}

To quantify how the concentration parameter and the MAH fitting parameter 
$\beta+\gamma$ affect the probability for a MW-mass halo to host MCs, 
we count the number of subhalos that can possibly host the MCs. 
Observational estimates suggest that the maximum circular velocity, $V_{\rm max}$, is $\sim50-60 \kms{}$ for the SMC 
\citep{Stanimirovic2004a,  Harris2006a} and $> 80 \kms$ for the LMC \citep{Olsen2011a, Kallivayalil2013a, van-der-Marel2014a}.
We choose halos that host at least two subhalos with $V_{\rm max}\geq55 \kms{}$ from the cosmological simulation. 
Overall, 28\% of the MW-mass halos in the simulation satisfy this criterion, 
broadly consistent with predictions of other simulations \citep{Boylan-Kolchin2010a, Busha2011a, 
Rodriguez-Puebla2013a, Cautun2014a}.
In the upper panels of Figure \ref{fig:fnsub_vcut}, we show the cumulative distribution 
functions of the concentration parameter $c$ 
and the MAH fitting parameter $\beta+\gamma$ of those halos selected from the cosmological simulation. 
In comparison, dashed lines show the same cumulative distribution functions for all the MW-mass hosts. 
The comparison clearly shows that, at a fixed mass, halos that host high-mass subhalos tend to have 
a lower concentration and more rapid late-time accretion. 
When selecting a MW host with both host mass and subhalo $V_{\rm max}$, the median concentration 
decreases to $c = 8.2$ as opposed to the median concentration $c=10.4$ for the entire simulated 
sample selected at a fixed mass. 
A Kolmogorov-Smirnov test on the distribution functions safely rejects the hypothesis 
that the distributions are the same with a $p$-value less than $6.9\times 10^{-18}$, confirming that the difference in the cumulative distribution function is statistically significant. 

In the lower panels of Figure \ref{fig:fnsub_vcut}, we show the fraction of the halos that host at least two 
subhalos with $V_{\rm max}\geq 55 \kms{}$ as a function of host concentration (lower-left), 
and as a function of $\beta+\gamma$ (lower-right). 
We split the entire halo sample into 10 bins according to their concentration or $\beta+\gamma$. 
The circles denote the mean fraction of such hosts for a given bin. 
The error bar for each bin represents the standard deviation estimated assuming the Poisson distribution. 
As one can see, the fraction of halos that host at least two $V_{\rm max}\geq55 \kms{}$ subhalos 
decreases with concentration and increases with $\beta+\gamma$. 
In this sample of a fixed halos mass, for host halos with concentration lower than 5, 
more than 67\% of them host two or more MC analogs. 
This probability drops below 20\% when the concentration is higher than 10. 
For the MAH fitting parameter $\beta+\gamma$, when $\beta+\gamma$ is higher than 1, 
more than half of those halos can host at least two MC-mass subhalos. 
The result shown in the lower panels of Figure \ref{fig:fnsub_vcut} demonstrates 
that the probability for a fixed mass halo to host MCs strongly depends 
on the concentration parameter and its late-time accretion.
A MW-mass halo can have a high probability to host MCs if its concentration is low and it has 
rapid late accretion.

To explore how varying accretion history and concentration affects the satellite 
galaxy populations, we split the halo sample into two groups according to the location 
of a halo in the $c$--$(\beta+\gamma)$ diagram of Figure \ref{fig:haloparams} (Panel B). 
We use a division line, which is shown by a dashed line in the right panel of 
Figure \ref{fig:haloparams}, to split the halo samples into two groups. 
The upper-left part of the diagram is named ``Group 1'' (GP1), and the lower-right part is named ``Group 2'' (GP2).
Halos in Group 1 are less concentrated, have more rapid recent accretion, and tend to host high-mass subhalos, while halos in Group 2 are more 
concentrated, have slower recent accretion, and do not tend to host high-mass subhalos. 
The division defined here is not intended to be a physical classification, 
but merely to roughly split the host halo samples into 
two subsamples with equal size so that we can contrast the differences between 
them in the following studies. 
With a larger sample, one could split the host population into finer bins to better 
elucidate the trends we explore in this paper. 
For the current sample from zoom-in simulations, to which we will apply our SAM, 
we have 20 halos in Group 1, and 18 halos in Group 2.

For each group, we show the general trend of the MAH and the concentration 
of the host halos as a function of redshift in Panels A and B of 
Figure \ref{fig:halo_groups}, respectively.  
We show the median and the 20\%, 50\%, and 80\% percentile of the distributions 
of these functions predicted by the cosmological simulation.
In the comparisons, the Group 1 hosts show relatively more rapid mass accretion 
at late times ($z<1$), while the Group 2 hosts have flat mass accretion histories at late times. 
The concentration also shows a clear difference between the two groups. 
The median values of the concentration for both groups stay around 5 from high redshift until $z\sim1$ 
for both halo groups. 
Group 1 halos slowly increase their concentration, with the $80^{\rm th}$ percentile being around 
5--10 at $z=0$. 
In contrast, Group 2 halos increase their concentration rapidly at $z<1$. 
At $z=0$, the 80 percentile of the concentration distribution for Group 2 hosts is between 10 and 15. 

In our simulated halo samples, the median concentration is 8.2 for Group 1 hosts and 12.5 for Group 2 hosts. 
The typical values of the concentration parameter of the simulated halos, especially those in Group 1, appear to be much lower 
than some values derived from fitting an NFW profile to observational kinematic data of the MW \citep[e.g.,][]{Battaglia2005a, Deason2012a}. 
One important difference to note is that the concentration values we report here are for halos in dark matter-only simulations, 
while the MW in reality has baryons. 
To make a fair comparison with existing kinematic constraints, we predict the circular velocity as a function 
of radius for the simulated halos by employing a toy model, which accounts for the gravitational potential of the disk and the bulge of the MW and the effect of halo contraction as a response to the formation of the MW galaxy.
In this toy model, we assume that the mass of the MW disk is $M_{\rm disk}=5.2\times 10^{10}\,\msun$, 
and the mass of the MW bulge is $M_{\rm bulge}=9\times 10^{9}\,\msun$ \citep{Licquia2015a}. 
We also assume that the disk follows an exponential profile with $r_{\rm d}=3$~kpc 
and that the bulge follows a Hernquist profile \citep{Hernquist1990a} with a scale radius $r_{\rm b}=0.6$~kpc \citep{Smith2007a}. 
We adopt the adiabatic contraction model proposed by \citet{Blumenthal1986a} to numerically compute 
the final dark matter density profile, ignoring other detailed dynamical effects, 
which can affect the circular velocity at about 5\% level \citep{Gnedin2004a, Choi2006a}. 

Panel C of Figure \ref{fig:halo_groups} shows the predictions for each halo group. 
The color bands and the central solid lines show the 20\%, 50\% and 80\% percentiles of the distribution and the median
of the predictions for each halo group. 
To show the effect of the baryonic components, we also plot in dotted lines the rotation curves resulted from only the dark matter halo with an NFW profile with the median value of concentration for each group.
In addition, the dashed line shows the rotation curves for the case where the MW disk and bulge are added, but the halo contraction effect is ignored.
By comparing these predictions, we find that the gravitational effect of the baryonic components of the MW substantially boosts the circular velocity in the inner halo $r<60$~kpc.
We overplot observational constraints from \citet{McMillan2010a}, \citet{McMillan2011a}, 
\citet{Wilkinson1999a}, \citet{Xue2008a}, \citet{Gnedin2010a}, and \citet{Watkins2010a}
on the same panel to compare with the predictions.
This comparison shows that, when the baryonic component and the contraction effect are taken into account, both halo groups, even with $c\sim8$ as predicted by dark matter only simulations, are consistent with most of the current kinematic data constraints, although only Group 2 is consistent with the \citet{McMillan2011a} constraint.

We also show the distribution of the cumulative subhalo mass function of each host group in Panel D of 
Figure \ref{fig:halo_groups}.
The mass function for Group 1 hosts is systematically shifted to higher masses relative to the mass function of Group 2 hosts. 
This is because the first two most massive subhalos in Group 1 hosts typically have higher 
mass than those of Group 2 hosts.  
The median mass of the most massive subhalo in Group 1 hosts is about $10^{10.75}\msun$, 
and $10^{10.25}\msun$ for Group 2 hosts. 
For this reason, the cumulative mass function of Group 2 hosts appears to be steeper 
than that of the Group 1 hosts in the regime of $N(>M_{\rm sub})<3$. 
We shift down the median of the cumulative mass function of Group 1 hosts by two 
and plot it as a red dashed line in the same figure, demonstrating that the two mass functions (with one of them being shifted) agree very well for $N\geq3$. 
This comparison indicates that the main difference in the subhalo population 
between hosts with different MAH and concentration is the masses of the two 
most massive subhalos. 
When those rare high-mass subhalos are excluded, the rest of the subhalo population is basically the same from halo to halo in terms of the mass distribution. 
We verify this statement for other properties of the satellite population when we apply our SAM on the host halos in the following sections.

\section{Model inferences from the MW satellite mass function}\label{sec:inference}

\subsection{Satellite galaxy stellar mass function}\label{sec:smf}

\begin{figure*}[htb!]
\begin{center}
\includegraphics[width=0.95\textwidth]{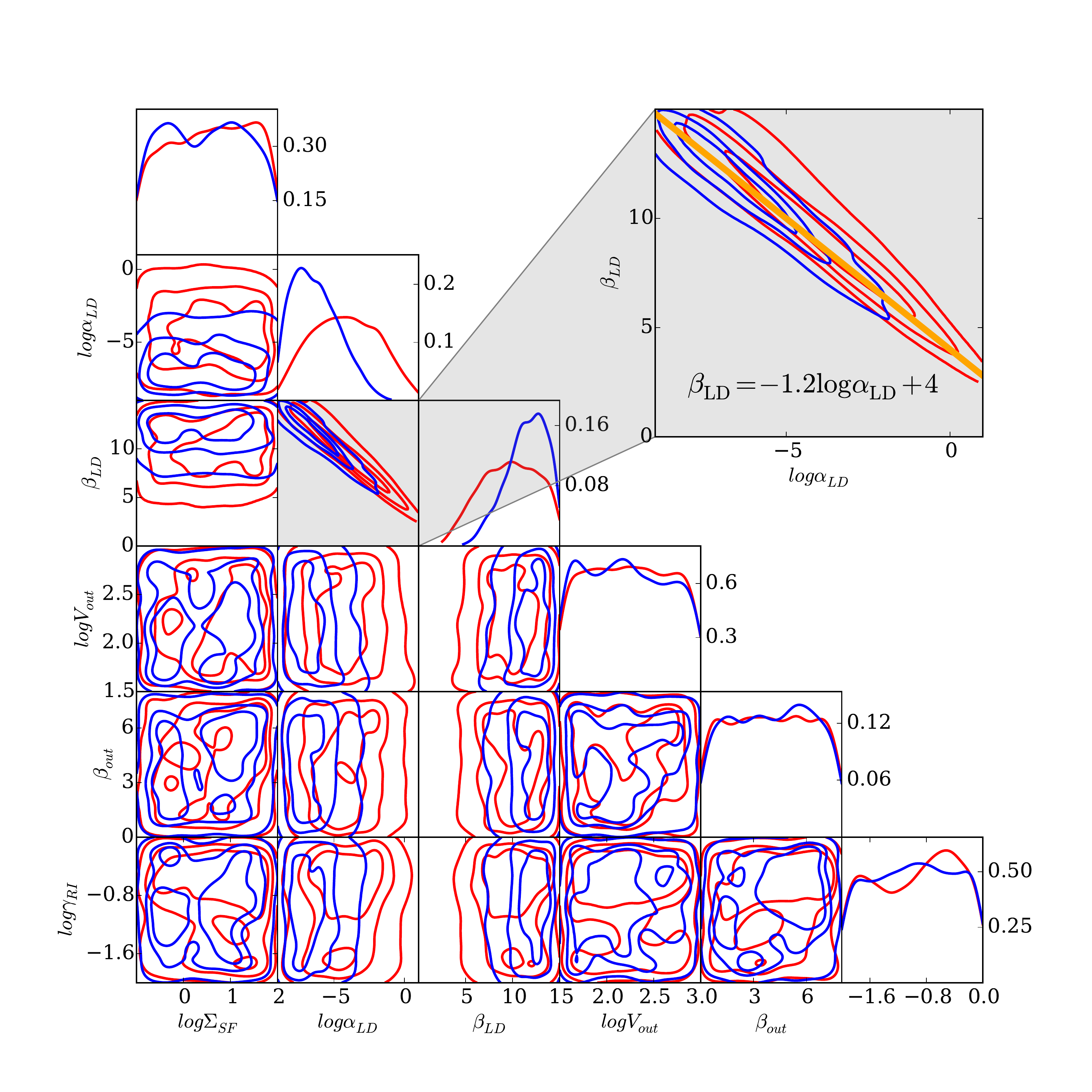}
\caption{Posterior probability distributions of the SAM parameters constrained with 
the MW satellite stellar mass function. 
The red contours denote the posterior distribution derived using the Group 1 halo prior, 
and the blue contours denote the distribution using the Group 2 halo prior.
The upper-right inset is an enlarged version of the posterior for the 
feedback outflow mass-loading parameters. 
The solid line in the inset denotes a linear relation between the two parameters 
as $\beta_{\rm LD} = -1.2 \log \alpha_{\rm LD} + 4$, which captures 
the degeneracy between the two model parameters in both runs. 
The figure shows that the inference of galaxy formation physics is strongly affected 
by the halo prior adopted. 
}
\label{fig:post}
\end{center}
\end{figure*}


\begin{table*}[htb!]
\begin{center}
\caption{Summary of semi-analytic model parameters} 
\centering
\begin{tabular}{c p{8cm} c c c}
\hline
\hline
\\ [0.ex]
notation & meaning of parameter & prior & best fit for GP1 & best fit for GP2
\\ [0.5ex]
\hline
$\Sigma_{\rm SF}$ & 
the threshold gas surface density for star formation in units of $\msun {\rm pc}^{-2}$  & 
$\log\Sigma_{\rm SF}\in[-1.0,2.0]$& 
no constraint &
no constraint
\\

$\alpha_{\rm LD}$ &
the normalization of the mass-loading factor for feedback outflow&
$\log\alpha_{\rm LD}\in [-9.0, 1.0]$&
-0.98 &
-5.85
\\

$\beta_{\rm LD}$ &
the power index for the circular velocity dependence of the mass-loading factor &
$\beta_{\rm LD} \in [0.0, 15.0]$ &
5.32 &
11.7
\\

$V_{\rm out}$ &
characteristic halo circular velocity in $\kms$, below which all outflow mass leaves the host halo &
$\log V_{\rm out} \in [1.5,3.0]$&
no constraint &
no constraint
\\

$\beta_{\rm out}$ & 
the steepness of the transition from total outflow for halos with $V_c<V_{\rm out}$ to no outflow for 
halos with $V_c>V_{\rm out}$ &
$\beta_{\rm out} \in [0.0, 8.0]$&
no constraint &
no constraint
\\

$\gamma_{\rm RI}$ &
the fraction of outflow mass reincorporated back into the halo &
$\log \gamma_{\rm RI} \in [-2.0,0.0]$&
no constraint &
no constraint
\\

$s_{\rm I}$ &
the fractional intrinsic scatter of the satellite mass function &
$s_{\rm I} \in [0.05,0.5]$&
no constraint &
no constraint
\\

\hline 
\hline
\end{tabular}
\label{tab:model}
\end{center}
\end{table*}
We have shown that, at a fixed halo mass, the formation history and concentration vary and span a wide range. 
Correlated with those variations, the subhalo population also varies.
We have split the simulated MW halos into two subsamples with roughly equal sizes 
based on their concentration and formation history to represent two different halo 
priors for a given final mass.  
In this section, we test the effects of halo priors defined by the two halo groups 
on the inference of galaxy formation model parameters by performing a Bayesian inference from 
the MW satellite stellar mass function.

The data we use in this paper is the stellar mass function of the MW satellite galaxies, 
down to the $11^{\rm th}$ most massive satellite galaxy ($M_*= 2.9\times10^5\msun$). 
We adopt the stellar masses and memberships of MW satellite galaxies compiled in 
\citet{McConnachie2012a}. 
\citet{Tollerud2008a} have shown that the incompleteness of the MW satellite galaxy count 
becomes important only for fainter dwarfs $M_{\rm v}>-7$ or $L<10^{5}L_{\odot}$ unless there is a significant low-surface-brightness population of satellites like Crater 2 \citep{Torrealba2016a}. 
We restrict this study to higher masses to avoid uncertainties in incompleteness corrections. 

The theoretical prediction for the mass function is straightforward, but a likelihood 
function of the satellite mass function for the data given model is needed to perform a Bayesian inference. 
Supported by the tests presented in Appendix \ref{app:likelihood}, we adopt the Negative Binomial 
Distribution (NBD) as a model for the likelihood function of the satellite stellar mass function, 
\begin{equation}
L(D|\theta) = \Pi_{i} P(N_i | r, p_i) = \Pi_i \frac{ \Gamma(N_i+r)} {\Gamma(r) \Gamma(N_i+1) } p_i^r (1-p_i)^{N_i}\,,
\end{equation}
where $N_i$ is the observed number of satellite galaxies for a given stellar mass bin $i$ per MW halo; 
the two parameters $r$ and $p_i$ are determined by the model as $r= { 1 }/{ s^2_{\rm I}}$, 
$p_i={1}/(1 + s^2_{\rm I} \mu_i)$; $\mu_i$ is the expected number for the $i^{\rm th}$ mass bin 
predicted by the model. $s_{\rm I}$ is the fractional scatter from the intrinsic scatter, $\sigma_{{\rm I},i}$,   
with respect to the Poisson scatter, $\mu_i$, defined as $s_{\rm I}\equiv \sigma_{{\rm I},i}/\mu_i$ (see Appendix \ref{app:likelihood}). 
We note that the value of $s_{\rm I}$ may vary as a function of mass bin and can be 
simulated for any given model if a large number of merger trees are utilized. 
In this paper, however, we assume it is a constant for the mass bins and 
treat it as a nuisance parameter to be sampled 
with MCMC because the number of the simulated hosts is not large enough to yield an accurate 
estimate for this parameter. 
We marginalize the posterior distribution of this parameter in the analysis.

We use MCMC to sample the posterior probability density distribution. 
Two separate runs are performed with the Group 1 hosts and the Group 2 hosts, respectively. 
In both runs, we allow six model parameters and the nuisance parameter $s_{\rm I}$ to 
vary within their prior ranges. 
A brief description of the model parameters and priors are listed in Table \ref{tab:model}. 
For a detailed explanation of these parameters, readers are referred to \citet{Lu2014b}. 
For each halo prior, we run the MCMC for 20,000 iterations with 144
parallel chains using the differential evolution algorithm
\citep{Ter-Braak2006a}. The convergence test is done with
the Gelman-Rubin test \citep{Gelman1992a}, requiring the potential scale reduction factor $\hat{R}<1.2$.
After removing outliers and pre-burn-in states, 
we obtain $\sim2,000,000$ posterior samples from the MCMC for each run. 

We show the 2-D marginalized posterior distribution for the Group 1 halo prior 
and the Group 2 halo prior in Figure \ref{fig:post}. 
We also list the ``best-fit'' model (maximum likelihood) parameters for each group 
in Table \ref{tab:model} for comparison. 
As one can see, under the same data constraints, using a different halo prior results 
in different posteriors of the model parameters. 
An obvious change is the normalization ($\alpha_{\rm LD}$) and the power index ($\beta_{\rm LD}$) for 
the parameterization of the mass-loading factor of outflow. 
In the model, the outflow mass-loading factor is parameterized as 
\begin{equation}\label{equ:massloading}
\eta = \alpha_{\rm LD} \left( \frac{V_{\rm c}}{220 \kms} \right)^{-\beta_{\rm LD}}\\,
\end{equation}
where $V_{\rm c}$ is the circular velocity of a halo, $\alpha_{\rm LD}$ and $\beta_{\rm LD}$ 
are model parameters to be constrained. 
The normalization parameter $\alpha_{\rm LD}$ for Group 1 is higher than that for Group 2. 
Correspondingly, the power index for Group 1 is lower than that for Group 2. 
The normalization for the mass-loading factor is defined as the mass-loading factor 
for halos with a circular velocity $V_{\rm c}=220 \kms{}$. 
A higher $\alpha$ requires a higher fraction of supernova energy to power outflows to keep 
the baryon fraction of a halo low. 
The Group 1 halos are those having high probability to host massive subhalos. 
The high-mass subhalos typically have higher baryon mass to start with 
before being accreted into the host. 
To keep the baryon mass fraction low in those high-mass subhalos, the model is 
required to have stronger feedback to suppress star formation. 
The Group 2 halos, which do not tend to have massive subhalos, typically require 
lower feedback to allow relatively higher stellar  mass fraction to fit the satellite 
mass function. 

As shown in Figure \ref{fig:post}, 
it is clear that the two parameters for the outflow mass-loading factor are strongly 
degenerate, regardless of which halo prior is adopted. 
The degeneracy can be approximately described by a linear function as 
$\beta_{\rm LD}= A\log \alpha_{\rm LD} + B$, where $A$ and $B$ are the slope and 
the intercept of the linear function. 
The straight orange line in the insert panel of Figure \ref{fig:post} shows this linear function  
with $A=-1.2$ and $B=4$. 
Recall that the outflow mass-loading factor is parameterized as Eq.~\eqref{equ:massloading}. 
Hence, if $\alpha_{\rm LD}$ and $\beta_{\rm LD}$ follow the aforementioned linear relationship, then we have 
$\log V_{220} = {1 / A}$ and $\log \eta = - {B / A}$. 
Using the values of $A$ and $B$ we derived from the degeneracy, we find that the outflow mass-loading factor $\eta\approx 2100$, when $V_{\rm c} = 32 \kms{}$. 
This is the generic feature for all models that are on the ridge of the degeneracy. 
This velocity scale is interesting because it is between the circular velocity of SMC 
and less massive satellites, such as Canis Major and Sagittarius dSph. 
The degenerate models collectively require extremely high outflow mass-loading 
factor for subhalos with circular velocity $\leq$$32 \kms$. 
It would be interesting to investigate the significance of this velocity scale using 
other data sets of galaxy populations.

\begin{figure*}[htb!]
\begin{center}
\includegraphics[width=0.85\textwidth]{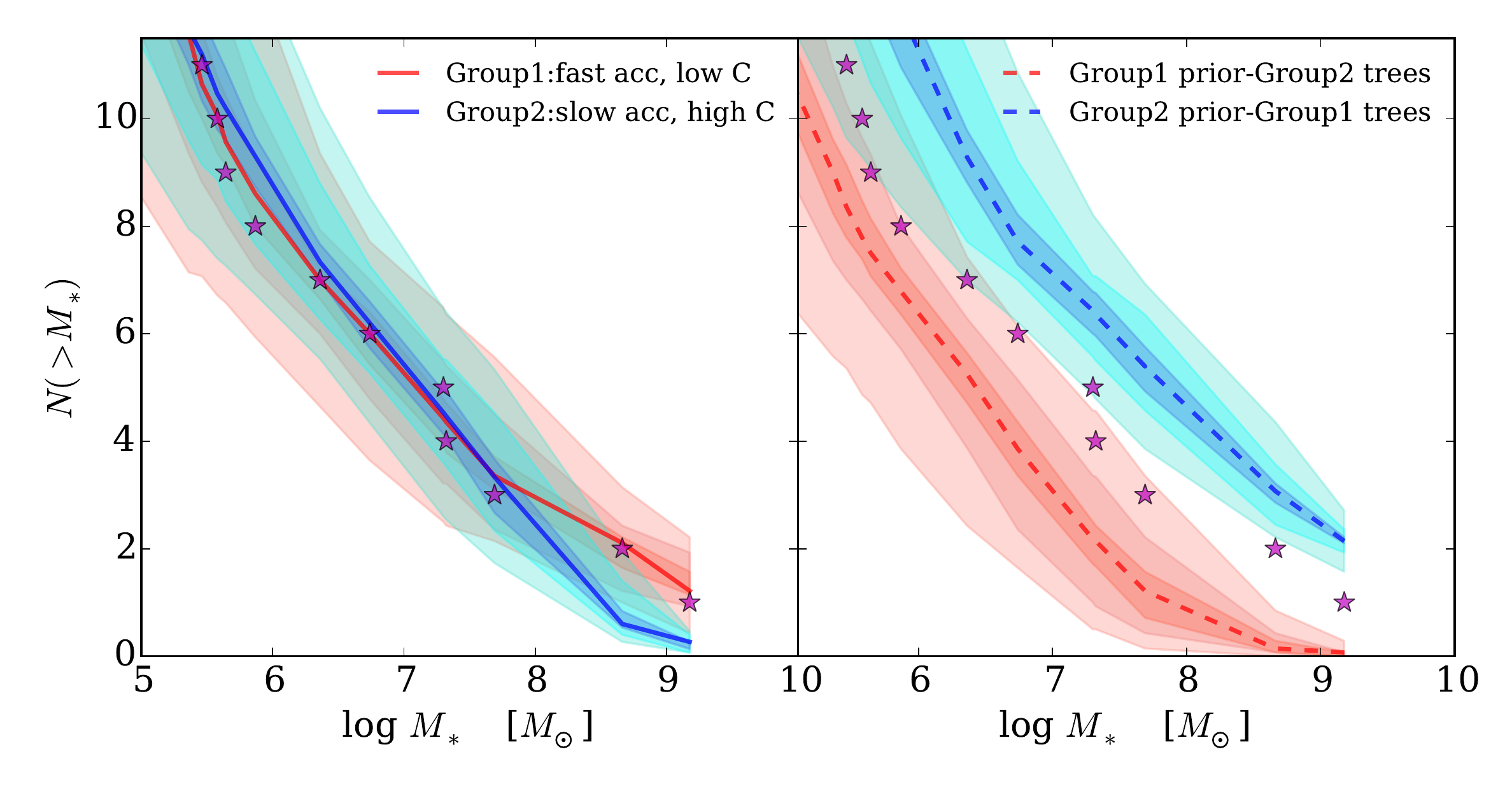}
\caption{{\bf Left}: The posterior predictive distribution of the MW satellite stellar mass function. 
The red is predicted by the model constrained using Group 1 halo prior, 
and the blue is predicted by the model constrained using Group 2 halo prior. 
The bands with decreasing intensity encompass the 20\%, 50\%, and 80\% predictive distribution 
for each model. 
Observational data denoted by star symbols are from \citet{McConnachie2012a}.
The model can match the stellar mass function using either halo priors. By eye, Group 1 halo prior 
results in better match to the data for MCs. 
{\bf Right}: The same posterior predictive distributions as the left panel but using \emph{mismatched} halo prior 
for each model. 
The red is predicted by the model constrained using Group 1 halo prior but applied to Group 2 halo merger trees. 
The blue is predicted using the model constrained using Group 2 halo prior but applied to Group 1 halo merger trees. 
Using mismatched halo priors, neither of the well-calibrated models fit the data.
}
\label{fig:mfunc_post}
\end{center}
\end{figure*}

We marginalize over the posterior to show how the models, together with their predetermined 
halo priors, reproduce the MW satellite stellar mass function. 
The reproduced stellar mass function using the two underlying dark matter 
halo priors are shown in the left panel of Figure \ref{fig:mfunc_post}. 
The red bands denote the posterior predictive distribution using the Group 1 halo prior. 
The solid line in the middle of the bands denotes the median of the predictive distribution, 
and the bands from darker to lighter color enclose 20\%, 50\%, and 80\% of the predictive 
probability distribution. 
Similarly, the blue bands show the predictive distribution using the Group 2 halo 
prior with the same three levels of confidence range.  
In addition, we also show the observed satellite mass function of the MW in the same figure. 
We find that when the model is applied to Group 1 hosts, it can achieve a very good match to 
the observed stellar mass function for the most massive 11 satellite galaxies. 
When the model is applied to Group 2 hosts, while the model matches the mass function equally 
well as using Group 1 hosts for satellites with mass lower than the SMC, it still tends to 
predict lower stellar masses for the two most massive satellites. 
It is because the hosts do not host high-mass subhalos, precluding solutions 
that can make sufficiently high mass galaxies to match the stellar mass of MCs. 
We compute the Bayes Factor, which is defined as the ratio of the marginalized likelihoods
of the models based on the Group 1 and Group 2 halo priors. We find that the Bayes Factor 
${\cal B} = {{\cal M}_1 / {\cal M}_2} = 1.42$, which indicates that the halo prior of 
Group 1 type MW host is only weakly preferred by the data over the halo prior 
of Group 2 type host. 
Due to large uncertainties, the model is not able to rule out either of the two halo 
priors using the observed MW satellite galaxy stellar mass function.

Using the constrained models, we show the biases in the model predicted MW satellite 
mass function if a mismatched halo prior is adopted. 
To illustrate the extreme case, we apply the model constrained using Group 1 halos 
to the merger trees of Group 2 halos, and the model constrained using Group 2 halos 
to the Group 1 halo merger trees to predict the satellite galaxy stellar mass function. 
The posterior predictive distributions are shown in the right panel of Figure \ref{fig:mfunc_post}. 
The model constrained for Group 1 halos predicts a significantly lower stellar mass function 
when Group 2 halo merger trees are employed. 
On the other hand, the model constrained for Group 2 halos predicts a too high mass function
when Group 1 halo merger trees are applied. 
The observed stellar mass function is excluded from the 50\% confidence range of 
the predictive distributions of both models. 
The MCs are even outside the 80\% confidence range. 
This result again demonstrates the strong influence of the underlying dark matter halo 
formation history on the hosted galaxies.

\subsection{Satellite-subhalo matching}

\begin{figure}[htb!]
\begin{center}
\includegraphics[width=0.45\textwidth]{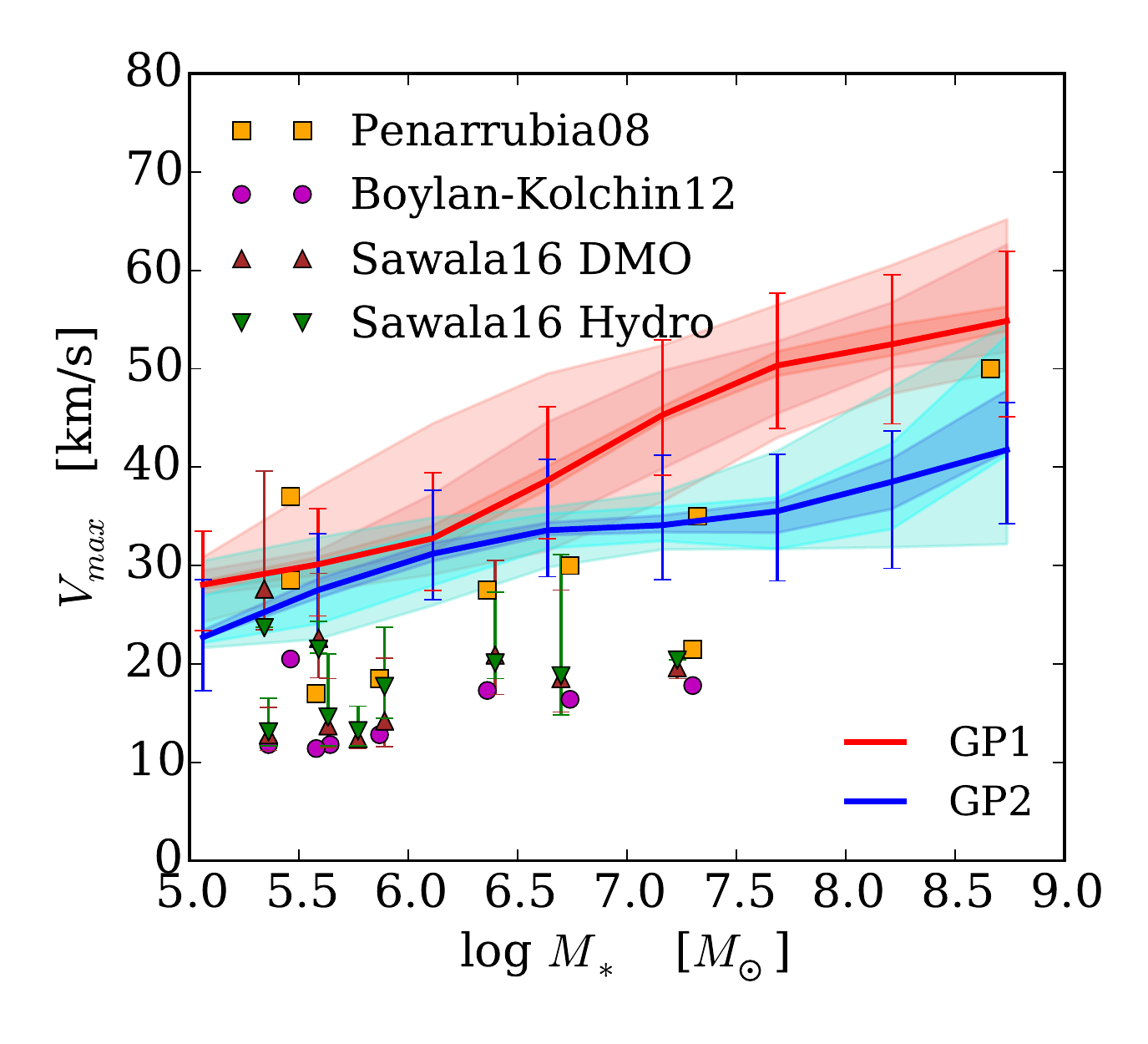}
\caption{
The posterior predictive distribution of the maximum circular velocity $V_{\rm max}$ of a 
subhalo as a function of stellar mass. 
The red and blue lines are predictions from the models constrained by the Group 1 and Group 2 halo priors, respectively. 
The bands with decreasing intensity encompass the 20\%, 50\%, and 80\% of the predictive distribution for each model.
The error bars show the averaged $1\sigma$ galaxy-to-galaxy scatter predicted by each model. 
The symbols denote the estimated $V_{\rm max}$ for MW classic dwarfs by \citet{Penarrubia2008a}, 
\citet{Boylan-Kolchin2012a}, and \citet{Sawala2016a}.
The formation history and concentration of a MW-mass host halo significantly affects the $V_{\rm max}-M_*$ relation of its satellites 
with $M_*\geq10^{6.5}\msun$. 
}
\label{fig:vmax}
\end{center}
\end{figure}

Using the posteriors, we can study the connections between MW satellite galaxies and 
their host subhalos based on the two halo priors and the corresponding models specifically 
tuned for each halo prior. 
Instead of choosing one single model (the best or any arbitrary one), we use the posterior 
samples obtained with MCMC to produce the posterior predictive distribution \citep{Lu2012a}. 
The predictive distribution marginalizes over the model uncertainties that are allowed by the 
uncertainties of the constraining data and quantifies the confidence levels of the predictions 
for the given model and prior. 
Figure \ref{fig:vmax} shows the maximum circular velocity, $V_{\rm max}$, of subhalos as a function of their stellar mass, as predicted by the models using the two halo priors.
The color bands with decreasing intensity denote the 20\%, 50\%, and 80\% confidence ranges 
of the predictive distribution for each model, and the solid lines show the median predictions. 
The error bars on each predictions represent the $1\sigma$ galaxy-to-galaxy scatter 
averaged over all the posterior sample models.
At high mass, $\log M_*/\msun >6.5$, the model based on the Group 1 halo prior 
predicts higher $V_{\rm max}$ than the model based on Group 2 halos for given stellar masses. 
This is because Group 1 hosts typically have higher mass subhalos than Group 2 hosts in this regime.
To fit the same satellite stellar masses, the model for Group 1 hosts has to populate satellite galaxies into the few relatively higher mass subhalos.
At lower mass, $\log M_*/\msun < 6.5$, the $V_{\rm max}-M_*$ relation predicted by 
the two models becomes similar.
This stellar mass scale corresponds to $V_{\rm max}\sim 30 \kms{}$ and the halo mass is about $3\times 10^9\msun$.
The $V_{\rm max}-M_*$ relation converges at this mass because this is where the subhalo mass functions of the two host groups become similar, as Figure \ref{fig:halo_groups} showed.
This is also reflected in the degenerate behavior of the feedback parameters as we discussed in 
\S\ref{sec:inference}. 
The models based on different halo priors require different strengths of feedback for higher-mass halos, but similar outflow mass-loading factors for halos with circular velocity lower than $\sim 30 \kms{}$.
It is also worth pointing out that while the constrained models 
vary substantially across the parameter space, the monotonic trend between the predicted 
stellar mass and subhalo $V_{\rm max}$ is preserved. 
This means that while the detailed subhalo-galaxy matching may vary depending on the 
host halo prior and specific model adopted,  
the rank order of the satellites based on their stellar mass still generally follows the 
depth of the gravitational potential of the subhalos.

We also compare the model prediction of the $V_{\rm max}-M_*$ relation with that derived 
from observations by \citet{Penarrubia2008a}, \citet{Boylan-Kolchin2012a}, and \citet{Sawala2016a} 
for a number of MW classical dwarf spheroidals in Figure \ref{fig:vmax}. 
These widely cited results have quite different $V_{\rm max}$ estimates for the same galaxy. 
Comparing them with our model predictions, we stress the importance of 
accurate estimates of $V_{\rm max}$ for model inference.
Among the three different estimates, the \citet{Penarrubia2008a} results have significantly 
higher $V_{\rm max}$ for a given stellar masses and thus agree more with the model based on Group 2 halo prior.
As argued in \citet{Boylan-Kolchin2012a}, \citet{Penarrubia2008a} assumed the NFW density profile 
and the size--mass relation of field halos for subhalos, which result in overestimation of $V_{\rm max}$. 
We also find that the assigned NFW profiles in \citet{Penarrubia2008a} have significantly 
larger sizes than what we find in the SAM predictions.
The radii where the rotation curve peaks, $R_{\rm vmax}$, for the classic dwarfs derived by \citet{Penarrubia2008a} are typically as large as 3-5kpc, which are rare and are only relevant for the subhalos of the MCs in our simulations.
\citet{Boylan-Kolchin2012a} matched the MW dwarfs with simulated subhalos based on their 
abundances and found much lower $V_{\rm max}$ for given stellar mass. 
Using both dark matter only simulations and hydrodynamical simulations, \citet{Sawala2016a} 
also found lower $V_{\rm max}$ values.
These results are overplotted in Figure \ref{fig:vmax}, with symbols noted as ``DMO'' 
for dark matter only simulations and ``Hydro'' for hydrodynamical simulations.
Compared to these results, our models, regardless of which halo prior is adopted, predict 
much higher subhalo $V_{\rm max}$ values at a given stellar mass.
All the data points from \citet{Boylan-Kolchin2012a} and \citet{Sawala2016a} are below the lower bound of the 80\% confidence range of the predictions.
The higher $V_{\rm max}$ values in our model indicate that the simulated subhalos are too dense compared to the observed classic dwarfs, as \citet{Boylan-Kolchin2011a} showed (also known as the ``too-big-to-fail'' problem).
By varying the halo prior, we show that the variation in halo concentration and MAH for fixed halo mass results in large scatter in the subhalo $V_{\rm max}-M_*$ relation.
However, the effect of the increased scatter is only at high mass and not at stellar masses relevant to the too-big-to-fail problem.
Our results demonstrate that, at this halo mass, if no baryonic processes are included, the probability for CDM subhalos to be consistent with MW dwarf kinematics is low.
At a lower mass scale, this discrepancy will have even stronger statistical significance \citep{Jiang2015a}.

We stress, however, the importance of understanding the baryonic processes, because recent high-resolution baryonic simulations of MW-mass halos have shown promise in addressing the ``too-big-to-fail'' problem \citep{Brooks2014a, Sawala2016a, Wetzel2016a}, largely because (1) tidal effects with the MW-mass stellar disk reduce the number of surviving subhalos \citep[e.g.,][]{Zolotov2012a, Zhu2016a, Wetzel2016a} and (2) stellar feedback can reduce the inner densities of subhalos at these masses \citep[e.g.,][]{Read2005a, Pontzen2014a, Brook2015a, Chan2015a}.

\subsection{Metallicity relation of MW satellites}

\begin{figure}[htb!]
\begin{center}
\includegraphics[width=0.45\textwidth]{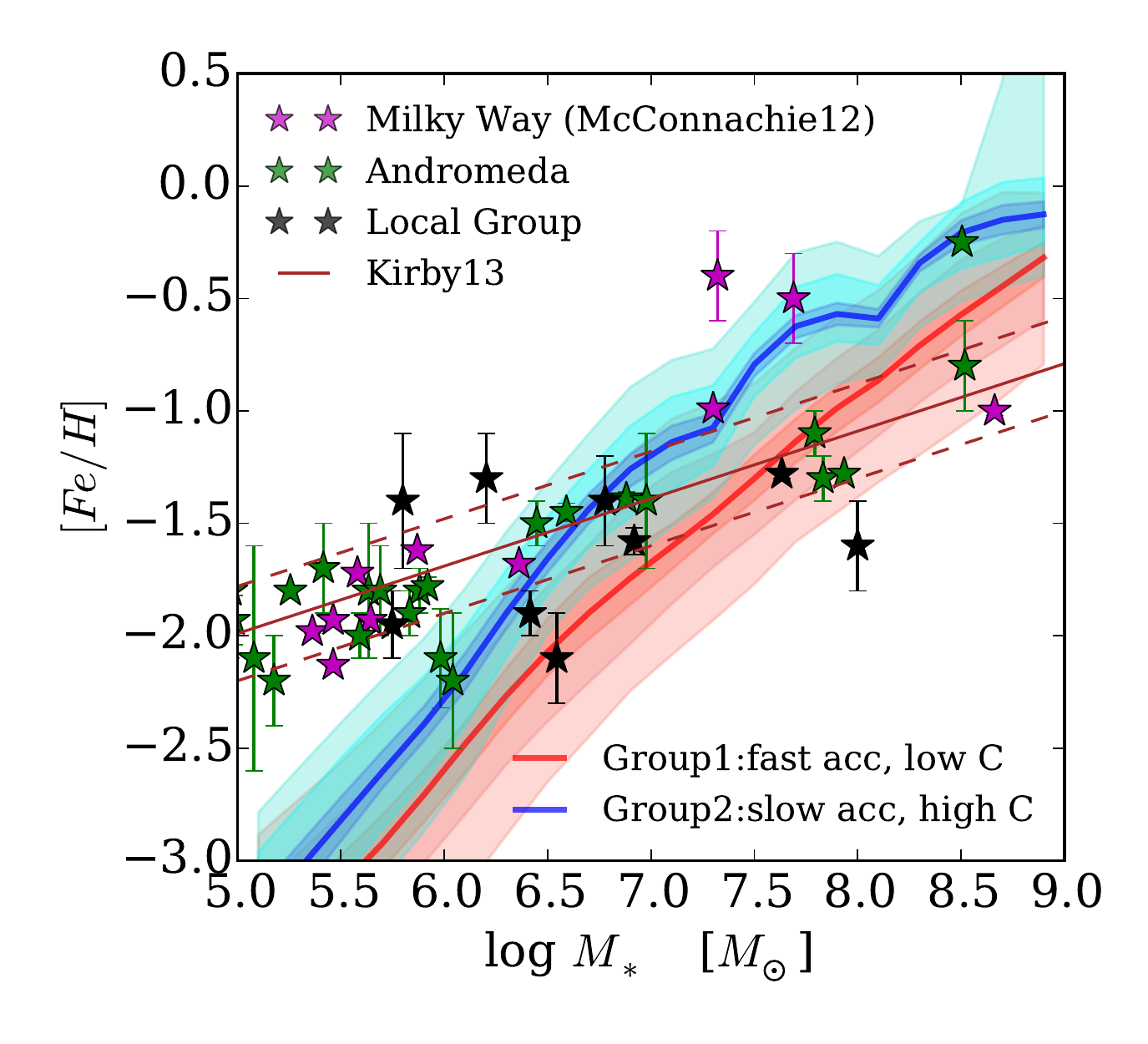}
\caption{The posterior predictive distribution of the stellar-phase metallicity as a function of 
stellar mass. 
The red and blue lines are predictions from the models constrained by the Group 1 and Group 2 halo priors, respectively.
The bands with decreasing intensity are the 20\%, 50\%, and 80\% predictive distribution for each model. 
Observational data compiled by \citet{McConnachie2012a} for different populations are shown by star symbols. 
The mean relation and its $1\sigma$ uncertainty derived by \citet{Kirby2013a} are shown as solid and dashed magenta lines.
The offset between the predictions shows that the formation history/concentration of a MW-mass halo significantly affects the inferred metallicity of its satellites in a semi-analytic model.
}
\label{fig:zstar}
\end{center}
\end{figure}

We now make predictions for the stellar-phase metallicity as a function of stellar mass for 
the MW satellite galaxies.
The posterior predictive distributions for both host halo priors are shown in Figure \ref{fig:zstar}. 
In the figure, the red bands are predicted using the posterior constrained for the Group 1 halo prior, 
and the blue bands are predictions using the posterior constrained with the Group 2 halo prior.
The bands with different intensity of each color show  the 20\%, 50\%, and 80\% predictive distribution. 
We find that while the model can be tuned to reproduce the satellite stellar mass function using either 
halo prior, they make rather different predictions for the stellar-phase metallicity-stellar mass relation. 
For the Group 2 prior, because the hosts typically do not have high-mass subhalos, the model needs weaker 
feedback and outflow to yield a relatively higher stellar mass for given subhalo mass. 
The consequence of the weaker outflow is to leave more metals in the galaxies, resulting in 
a higher metallicity for a given stellar mass.
In contrast, the Group 1 prior imposes host halos that typically host high-mass subhalos. 
The high-mass subhalos require stronger outflows, which eject more metals from the galaxy. 
The result suggests that for a given galaxy formation scenario, when it is calibrated to the stellar mass 
function, the properties of the underlying dark matter halo, including the formation history, concentration 
and subhalo masses, can leave imprints in the satellite galaxy metallicity relation. 
Because the high-mass end of the subhalo mass function is correlated with the formation history 
and concentration of the host halo, the metallicity may also reflect the formation history and other properties of the 
MW host halo and its analogues. 

The stars with error bars are observational data compiled by \citet{McConnachie2012a}, and the mean observational relation is shown as solid and dashed magenta lines \citep{Kirby2013a}.  We note that our mass-metallicity relation using either halo prior has a steeper slope than the observed relation.  Using this relation to constrain our galaxy formation models is beyond the scope of this paper, but the difference in the relations using the different priors may provide useful comparisons with data in future work.

\subsection{Satellites accreted with LMC}

In a CDM cosmology, the higher amplitude of the power spectrum at small scales gives rise to a 
large number of substructures in a MW-mass halo \citep[e.g.,][]{Klypin1999a, Moore1999a, Kravtsov2004b}. 
\citet{Wetzel2015a} studied the fraction of satellites of MW-mass halos 
accreted as part of a group in the ELVIS simulation \citep{Garrison-Kimmel2014a} \footnote{\href{http://localgroup.ps.uci.edu/elvis}{http://localgroup.ps.uci.edu/elvis}} and found that higher mass satellite galaxies 
have higher probability of hosting satellites when they are finally accreted into MW. 
Using a similar method for tracking the accretion history of subhalos, we explore the number of 
satellites galaxies that are accreted into the MW host halo together with the most massive subhalo.  
These smaller halos are subhalos of the subhalo of the MCs and are accreted into the MW 
host with the MCs. 
These sub-subhalos may remain gravitationally bound with 
the MC subhalo, or may have been tidally stripped from the MC subhalo potential and live in the 
vicinity of the MCs. 
Using the models that are calibrated to match the satellite mass function, we test if the satellite 
galaxies that are accreted with the MCs can provide information to discriminate the halo priors.

\begin{figure}[htb!]
\begin{center}
\includegraphics[width=0.45\textwidth]{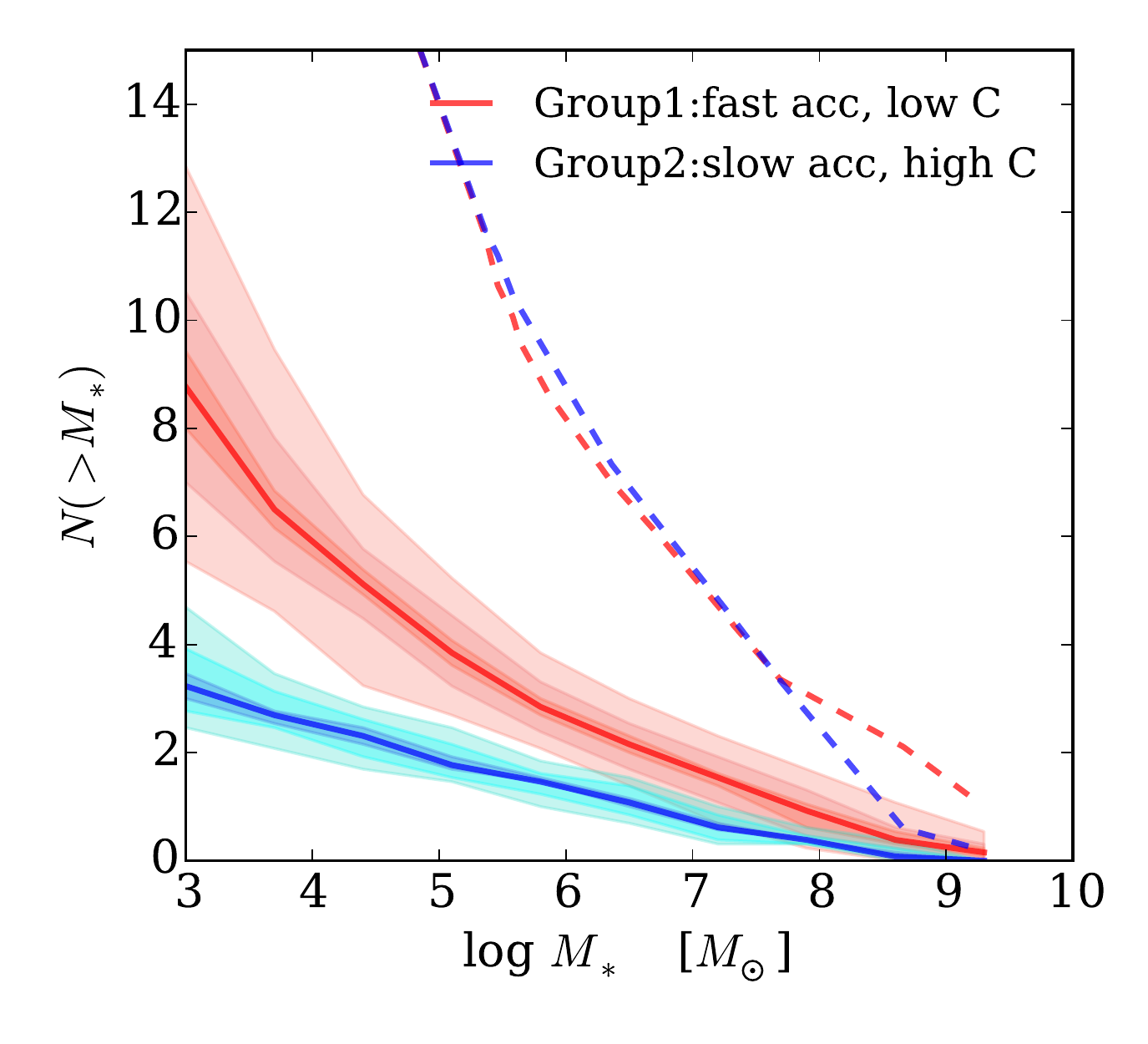}
\caption{The posterior predicted cumulative mass function of satellite galaxies that were accreted into the MW host halo as a satellite of the most massive subhalo, assumed to host an LMC-like satellite galaxy.
The solid red and blue color bands denote the predictions made by the models constrained by the Group 1 and Group 2 halo priors, respectively. 
The bands with decreasing intensity are the 20\%, 50\%, and 80\% predictive distribution for each model. 
For comparison, the dashed lines show the median stellar mass function of all satellite galaxies 
predicted by the respective models. 
}
\label{fig:sat}
\end{center}
\end{figure}

In the SAM prediction, we identify the most massive satellite galaxy in each MW-mass halo as its ``LMC'' and track the merger history of its subhalo using the merger tree. 
In the merger tree, we find the time when it is just accreted into the MW host and all subhalos 
hosted by the ``LMC'' host subhalo. 
We then track the history of all galaxies hosted by the subhalo since this epoch and 
identify the ones that are still present in the MW halo at $z=0$.
We repeat this calculation for every model in the posterior samples to produce a posterior 
predictive distribution.  
In Figure \ref{fig:sat}, we show the stellar mass function of the satellite galaxies that 
are accreted into the MW host together with the ``LMC'' as its satellites predicted by the constrained models 
at $z=0$. 
We show the predictions using the two different host halo priors. 
We find that Group 1 hosts have systematically higher numbers of satellites that are 
accreted with the most massive satellite than Group 2 hosts.
Down to $10^3\msun$, a Group 1 host typically has more than 8 satellite galaxies accreted 
with the most massive satellite, but a Group 2 host has no more than 5 satellite galaxies with stellar mass higher than $10^3\msun$ accreted with its ``LMC''.
The reason is simply that the subhalos of the ``LMC'' in Group 1 hosts are typically more massive 
than those in the Group 2 hosts, and thus they bring in more low-mass satellites.
While many of these small satellites have been stripped out of the ''LMC'' subhalo, they are still located near it in the simulation.
Transforming their positions into a coordinate system centered on the host halo, 
we find that on average 57\% of the small satellites are located within 40 degrees of the most massive satellite, and about 80\% of them are located within 100 degrees \citep[see also][for detailed analyses]{Deason2015a, Yozin2015a, Jethwa2016a, Sales2016a}.
In the year of 2015, 21 new dwarf galaxy candidates were discovered around the MW, and a significant fraction of these are within 50 degrees of either the LMC or the SMC~\citep[e.g.,][]{Drlica-Wagner2015a}. Although it is not yet clear how many of these candidates are true satellites of the MCs, our result suggests that identifying low-mass satellite galaxies that are associated with MCs may provide useful constraints on the subhalo mass of the MC and the MW host prior.

\subsection{Central galaxy stellar mass}

\begin{figure}[htb!]
\begin{center}
\includegraphics[width=0.45\textwidth]{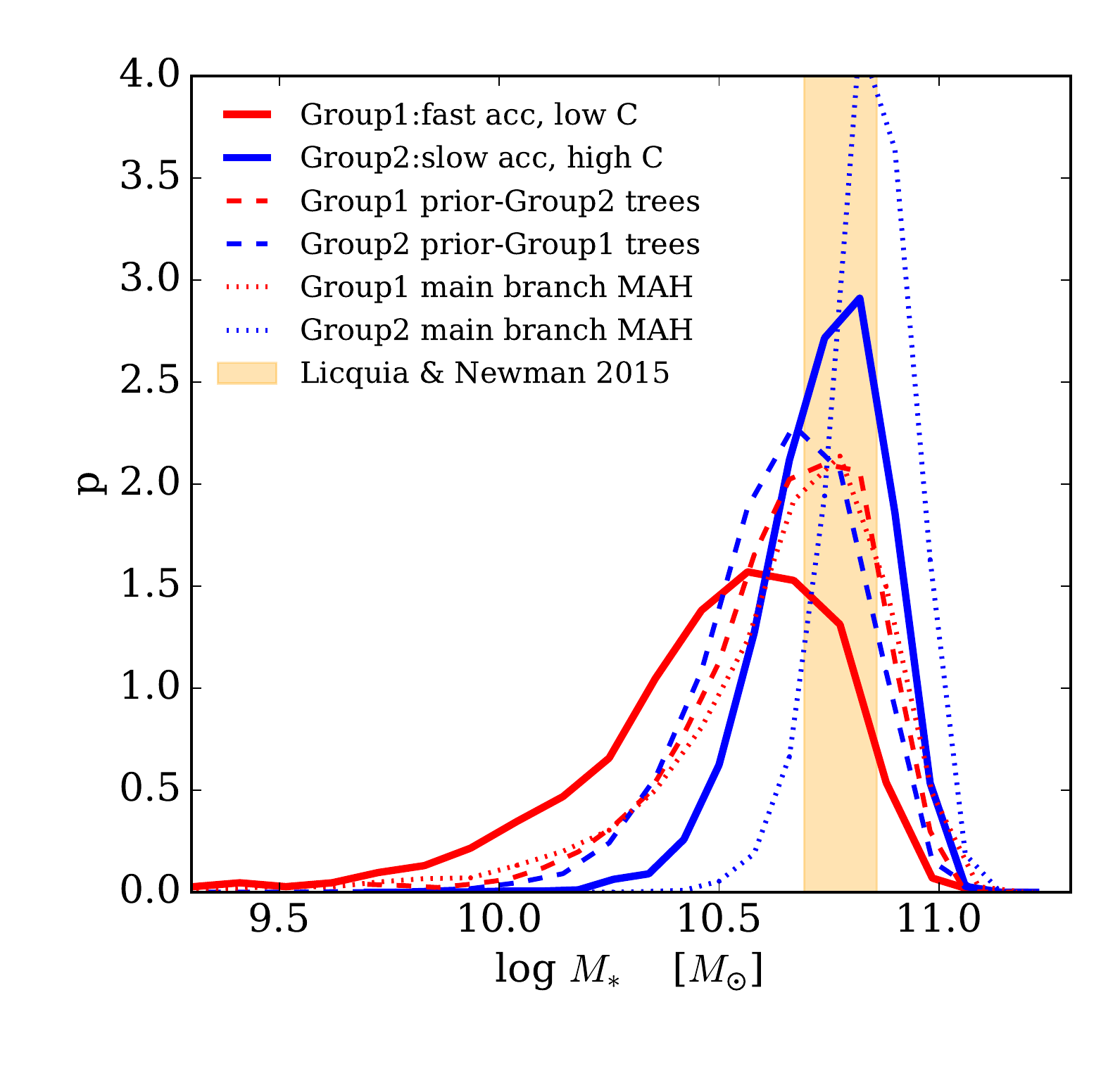}
\caption{
The posterior predicted probability distribution function for 
the central galaxy stellar mass of the MW host halos. 
The red solid line denotes the prediction of the model constrained using the Group 1 halo prior, 
and the blue solid line denotes the prediction of the model constrained using the Group 2 halo prior. 
The dashed lines show the same posterior predictions using the \emph{mismatched} halo prior for each model. 
The red dashed line is predicted by the model constrained using the Group 1 halo prior but applied to Group 2 halo merger trees. 
The blue dashed line is predicted using the model constrained using the Group 2 halo prior but applied to Group 1 halo merger trees. 
The dotted lines show the same predictions as the solid lines but using the main-branch MAHs of the host halos in each halo group.  
The orange band covers the observational estimate of the MW galaxy stellar mass ($M_*=(6.08\pm1.14)\times10^{10}\msun$) by \citet{Licquia2015a}. 
The model for the Group 1 halos, which have relatively lower concentration and more rapid recent accretion, predicts lower stellar mass for the central galaxy than the model for the Group 2 halo prior.
Models for both halo priors predict central galaxy stellar mass consistent with current observational constraints. 
}
\label{fig:central}
\end{center}
\end{figure}

The halo prior discussed in this paper is characterized by the concentration, MAH, and the high-mass subhalo content 
of the host halo. These host halo properties are expected to affect the mass of the central galaxy \citep[e.g.][]{Lehmann2015a, Zentner2016a}. 
In this subsection, we extrapolate the models constrained to the MW satellite stellar mass function for different halo priors to study 
the effect of the halo prior on the stellar mass of the MW galaxy. 
As we discussed in \S\ref{sec:sam}, the modeling scheme adopted for previous sections 
does not follow the formation of the central galaxy to focus computation on the satellite galaxy populations. 
In this section, we employ the full model \citep{Lu2011b}, which has the additional ability to make predictions for the central galaxy.
We run the full model with the same posterior parameter samples as in previous sections 
to predict the probability distribution of the central galaxy stellar mass that is implied by the constrained models. 
In Figure \ref{fig:central}, 
the red solid line shows the prediction of the constrained model using the Group 1 halo prior, 
and the blue solid line shows the prediction of the constrained model using the Group 2 halo prior. 
When both models are best constrained to the observed MW satellite stellar mass function, 
the two models predict different stellar masses for the central galaxy. As we show in the figure, the Group 2 halo prior 
systematically predicts higher central galaxy stellar masses than the Group 1 halo prior.

The difference in the predicted central galaxy stellar mass stems from two effects.
First, the halos of these two host halo groups have different formation histories and subhalo content. Second, the models that are best constrained to match the MW satellite galaxy stellar mass function for each halo prior are different.

By construction, Group 1 halos typically form later and host more high-mass subhalos than the Group 2 counterparts.
To show the effect of different halos on the central galaxy mass, we apply a fixed model to different merger trees to predict the central galaxy stellar masses. 
In Figure \ref{fig:central}, the dashed red line shows the central galaxy stellar mass distribution predicted by the model constrained using Group 1 halo prior but applied to Group 2 merger trees. Similarly, the blue dashed line shows the same prediction made by the model constrained using Group 2 halos but applied to Group 1 merger trees. We find that, for a given model, Group 1 halos systematically form lower mass central galaxies than Group 2 halos. 
The effect could stem from the fact that Group 1 halos host more high-mass subhalos, which can lock up a considerable fraction of baryons that would have been available to fuel star formation in the central galaxy.
However, we expect that this effect is relatively small, as even the most massive subhalo in Group 1 halos typically makes up a rather small fraction (typically $< 15\%$) of the total mass. 
To test if the difference in the central galaxy mass between the two halo groups still exists when the effect of locking baryons in subhalos is excluded, we apply the models to the main-branch halo MAHs of the merger trees, in which galaxy formation in subhalos is completely ignored as the baryons associated with the halo in this treatment can cool only into the central galaxy. 
The red dotted line in Figure \ref{fig:central} shows the predicted central galaxy stellar mass distribution of Group 1 halo main-branch MAHs, and the blue dotted line shows the prediction of Group 2 halo MAHs. 
One can find that while all central galaxies are predicted to have higher stellar masses than when we use the full merger trees (in this case, baryonic mass in halos is unable to cool and to form galaxies in subhalos), Group 2 halos still form systematically more massive central galaxies than Group 1 halos, even after removing the effect of locking up baryons in subhalos. 
In summary, Group 2 halos, which have higher concentrations and slower recent accretion, form higher stellar mass for the central galaxy than Group 1 halos mainly because Group 2 halos typically form earlier and are denser in the center. Star formation, at least in the model \citep[e.g.][]{Lu2014b}, is more efficient at higher redshift and denser halos. Consequently, the central galaxies of Group 2 halos form more stars over the history than those of Group 1 halos.

We also test the effect of different models.
As we demonstrated in \S\ref{sec:smf}, different levels of feedback are required by the two different halo priors to fit the MW satellite stellar mass function for each halo prior. 
Group 1 halos require a higher level of feedback than Group 2 halos. 
When the model is extrapolated to work on the MW host halo, 
the stronger feedback in Group 1 halos prevents stars from forming more efficiently than in Group 2 halos. 
We can see this effect by applying the two different models on the same merger tree set (comparing the blue dashed with red solid lines or the blue solid with red dashed lines). 
The blue dashed line in Figure \ref{fig:central} shows the prediction of the model constrained for the Group 2 halo prior but applied to Group 1 halos. The difference between this line and the red solid line shows the effect of changing galaxy formation models that are constrained for different halo priors. 
This is the same effect that can be seen by comparing the blue solid line and the red dashed line.

In Figure \ref{fig:central}, we also include the observational estimate of the MW galaxy stellar mass of \citet{Licquia2015a}, who constrained the stellar mass of the Milky Way galaxy to $M_*=(6.08\pm1.14)\times10^{10}\msun$. 
Both of the model predictions (the solid red and blue lines) agree with the observational result quite well. 
The Group 2 halo prior seems to agree with the observational estimate better, as its predicted stellar mass distribution peaks right at the observational constraint, but the Group 1 halo prior is also consistent with the observational result. 
We stress that the predictions for the central galaxies are made with models that 
are specifically constrained to the MW satellite galaxies. No information from higher-mass scales 
is used to constrain the models. 
The prediction could change if the physics that acts on the MW-sized galaxies is different from what we infer from the MW satellite galaxies. 
We also note that we have fixed the halo mass in this paper. The Group 1 halo prior would achieve a better match with the data if the host halo of the MW has higher mass than we have assumed here.

\section{Summary and Discussion}\label{sec:discussion}

In this paper, we have analyzed a suite of $N$-body cosmological simulations of 
MW-mass halos with various mass-assembly histories and concentrations.  
The analysis shows that these properties of a host halo are correlated with 
the probability for a halo to host high-mass subhalos. 
For fixed halo mass, halos that have rapid recent accretion, 
characterized by the fitting parameter $\beta+\gamma$ of the mass-assembly history, 
tend to host high-mass subhalos that still exist at the present day. 
These halos also tend to have a lower concentration. 
The result suggests that the existence of high-mass satellite galaxies may provide 
a clue to the formation history and the structure of the underlying host dark matter halo.  By applying a semi-analytic model to halos with different mass-assembly histories, we show herein that the best-fit galaxy formation models depend sensitively on the host mass-assembly history.  Therefore, it may be possible to infer the mass-assembly history of hosts, and the properties of hosts that correlate with the assembly history, based on properties of their satellite galaxies. 

In this work, we focused specifically on the MW and its satellite system, especially the unusually high mass satellites, the Magellanic Clouds. 
As suggested by recent observational estimates using different probes, the dynamical mass 
of the LMC is at least $1.7\times10^{10}\msun$ \citep{van-der-Marel2014a} or even higher 
\citep[e.g.,][]{Vera-Ciro2013a,Penarrubia2016a}. 
The high masses of the MCs have important implications for the structure of the MW's dark halo, as this would suggest that the host halo of the MW is more likely to have experienced
rapid accretion since $z=0.5\sim 1$ \citep[e.g.,][]{Besla2010a} and that it may have relatively lower 
concentration than typical halos at the same mass (i.e., Group 1). 

This argument of a low value for the MW halo concentration seems to be, at face value, in tension with some results in the literature.  Observations of kinematic tracers of the MW can provide direct constraints of the dynamical mass 
distribution of the MW halo. 
Some results suggest that the MW halo appears to have a concentration value
as high as or even higher than 18 \citep[e.g.,][]{Battaglia2005a, Deason2012a}.
One possible solution to the tension is that the MW halo is a significant outlier or an early-formed halo with the existence of the MCs as 
a ``transient'' coincidence ~\citep[e.g.,][]{Deason2016a}.
We stress, however, that the formation of a central galaxy in the halo center can cause a contraction of the dark matter matter 
in the inner halo. 
As we have shown in Section \ref{sec:mah}, it is still plausible that the MW has formed in a low-concentration halo, but  
the effects of the baryonic components of the MW give rise to 
a centrally peaked rotation curve, which can be mimicked by a halo with a much higher concentration value (although this may be in tension with \citet{binney2015a}, who argue against halo contraction).
Moreover, if the mass of the Milky Way host halo is higher than that assumed in our simulations, the low-concentration rapid-accretion hosts could achieve better fits to the Milky Way kinematic constraints \citep[e.g.][]{Wang2012b, Cautun2014b}.

These disparate conclusions about the MW halo's concentration can be solved with more data and better modeling of the dynamical effects of the Magellanic Clouds on the MW.  Our analysis suggests that more accurate measurements of the MW halo kinematics within $r<60$~kpc 
are still needed to tighten the constraints on the radial mass distribution of the MW halo. 
We also note that, with dark matter-only simulations, \citet{Busha2011b} found that 
MW analogs, when selected based on not only the virial mass but also the phase-space position of the MCs, tend to have slightly higher concentration values ($c\sim 11$).
It is worth noting, however, given the close distances of the MCs to the Galactic center, the observed speeds of the MCs are likely affected by the baryonic component of the MW Galaxy. Also, the observed speeds of the MCs can be biased if the MCs are not yet in equilibrium orbits. As such, more detailed modeling of the kinematics of the MCs is needed to understand what constraints on the host halo concentration can be derived.

The connection between the host halo properties and the subhalo populations also has interesting implications for the properties of the satellites of the MW and the stellar mass of the MW galaxy itself.  
To understand these implications, we constrained a flexible galaxy formation model with the observed MW 
stellar mass function of classical dwarf galaxies with different priors on the formation history and concentration of the host halo. 
We studied how different halo priors affect the 
inference of galaxy formation physics and, more interestingly, inferred what observations 
can help distinguish between different host halo priors and constrain the feedback processes.

First, we found that when assuming different host halo priors for the model, 
the observed stellar mass function requires a different strength of and circular velocity 
dependence for feedback (Figure \ref{sec:inference} and Table \ref{tab:model}). 
After the uncertainties in relevant processes of the model
are marginalized, we find that host halos with rapid recent accretion require stronger 
feedback than those without rapid recent accretion to fit the satellite galaxy mass function. 
The reason for this is because halos with rapid recent accretion 
tend to host high-mass subhalos, which then requires a lower stellar-mass-to-halo-mass ratio 
to match the observed mass function. 
Moreover, our model inference from the MW stellar mass function weakly prefers the host halo 
prior that the MW host halo had rapid recent accretion and has lower concentration than typical 
halos with a similar virial mass. 
However, due to large uncertainties in the model, especially the uncertainties in the feedback 
processes, our model inference cannot completely rule out the alternative prior that 
the MW host halo has slow recent accretion and higher concentration. 
This result suggests that, in order to determine the properties of 
the host halos using galaxy formation models, one has to better understand the feedback 
processes in galaxy formation.

Second, we found that to match the observed stellar mass function, 
different host halos require a significantly different mapping between subhalo mass or circular velocity 
and stellar mass, or $V_{\rm max}-M_*$ relation  (Figure \ref{fig:vmax}).  
The difference in the subhalo maximum circular velocity, $V_{\rm max}$, 
for a given stellar mass can be as large as $\sim 25 \kms{}$ between different simulated host halos. 
The scatter from the host halo prior, however, is not large enough to 
account for the observationally inferred low $V_{\rm max}$ values of the MW classical 
dwarfs \citep{Boylan-Kolchin2011a}, at least for the fixed halo mass employed in this paper. 
Moreover, this difference in the $V_{\rm max}-M_*$ relation between different simulated host halos 
mainly exists in satellites more massive than $M_*\sim10^{7}\msun$, 
and vanishes for satellites with mass lower than $\sim10^{6.5}\msun$. 
This is because the difference in the subhalo mass function of different host halos of a given mass 
only exists in the very high-mass end, where statistical fluctuation matters. 
When looking at the number counts of low-mass satellite galaxies, the MW 
is not a significant statistical outlier as it is for high-mass satellites \citep[e.g.,][]{Strigari2012a}. 
Ideally, a stronger test of the CDM cosmology with local group satellite galaxy kinematics 
should use very low-mass galaxies, in which the subhalo population suffers less from 
the statistical fluctuations inherent from the halo-to-halo scatter \citep{Purcell2012a, Jiang2015a}. 

Third, we find that the feedback processes assumed in the model strongly affect 
the mass--metallicity relation of dwarf galaxies (Figure \ref{fig:zstar}). 
This is because feedback not only suppresses star formation but also governs 
the flow of baryons, including metals, produced during galaxy formation \citep[e.g.,][]{Lu2015d, Lu2015b}. 
When a model requires stronger outflow to overcome the deeper gravitational potential of 
larger subhalos, it inevitably results in a lower metallicity in dwarf galaxies unless 
the metallicity of the outflow is different from that of the ISM \citep{Dalcanton2007a}. 
Therefore, one can find clues about the properties of the dark matter halo potential by 
modeling the strength and metallicity of galactic outflow.

Fourth, we find that halos with different MAHs, concentrations, and subhalo populations
predict different numbers of small satellite galaxies that are accreted together with the MCs. 
When constrained to reproduce the observed satellite galaxy stellar 
mass function, the host halos that have recent rapid accretion and lower concentration tend 
to host more satellites associated with the most massive satellite. 
The result suggests that finding and identifying low-mass satellite galaxies down to $M_*\sim10^3\msun$
that are physically or historically associated with high-mass satellites may provide useful constraints 
on the property of the host halo \citep[see also][]{Deason2015a}. 
The ongoing surveys for low-mass satellite galaxies within the MW and especially around the vicinity of the MCs 
\citep[e.g.,][]{Bechtol2015a, Drlica-Wagner2015a} may provide useful data for further 
theoretical investigations. 

Finally, assuming the model constrained to observations of MW satellite galaxies can be extrapolated to the MW host halo, we find that the host halo properties that are connected to the satellite galaxy populations also affect the formation of central galaxies.  
Since halos with lower concentration and more rapid recent accretion typically form later and require stronger feedback to match the satellite stellar mass function,
these halos form a central galaxy later and with lower stellar mass than halos with higher concentration and slower recent accretion. 
These effects explain the phenomenology of halo occupation and abundance matching models in which at fixed mass low-concentration halos are required to host a central galaxy with lower stellar masses \citep[e.g.][]{Reddick2013a, Lehmann2015a, Hearin2016a, Zentner2016a}, 
which is attributed to the effect of the assembly bias \citep{Gao2005a, Wechsler2006a, Jing2007a}.
While the effects of the assembly bias on galaxy populations have been studied with SAMs on large-scale survey data in previous works \citep[e.g.][]{Croton2007a, Wang2013a}, we have demonstrated in this paper that, at a fixed halo mass, the formation history of the host halo can affect both the central galaxy and the satellite populations simultaneously.

The interplay between mass-accretion history, host halo properties, and galaxy evolution has important consequences beyond the MW.  Interestingly, the Andromeda galaxy also has massive satellites, and thus may possibly have rapid recent accretion too.  
It would be interesting to investigate how probable it is that in CDM cosmology two nearby $L^*$ galaxies both host similar high-mass satellite galaxies, and how the accretion of high-mass subhalos correlates with the large-scale environment. 
These studies will shed further light on the position of our own galaxy, the Milky Way, in the cosmic web.

\acknowledgments
We thank Matthew Becker for providing the {\tt c125-2048} simulation, which was run using computing resources at SLAC. The resimulations were performed using computational resources at SLAC and at NERSC. We thank the SLAC computational team for their consistent support.  We also acknowledge the Ahmanson Foundation for providing computational resources used in this work.  Support for programs HST-AR-13270.003-A, HST-AR-13896.005-A and HST-AR-13896.009-A was provided by NASA through a grant from the Space Telescope Science Institute, which is operated by the Association of Universities for Research in Astronomy, Inc., under NASA contract NAS 5-26555.
Y.L. thanks Evan Kirby, Josh Simon, Robyn Sanderson, Shude Mao, and Simon White for useful discussion. 
S.T. was supported by the Alvin E. Nashman Fellowship in Theoretical Astrophysics.  A.R.W. was supported by a Moore Prize Fellowship through the Moore Center for Theoretical Cosmology and Physics at Caltech and by a Carnegie Fellowship in Theoretical Astrophysics at Carnegie Observatories.

\bibliographystyle{yahapj}
\bibliography{references}

\appendix

\section{Likelihood function for the Milky Way satellite galaxy stellar mass function}\label{app:likelihood}

\begin{figure*}[htb!]
\begin{center}
\includegraphics[width=0.9\textwidth]{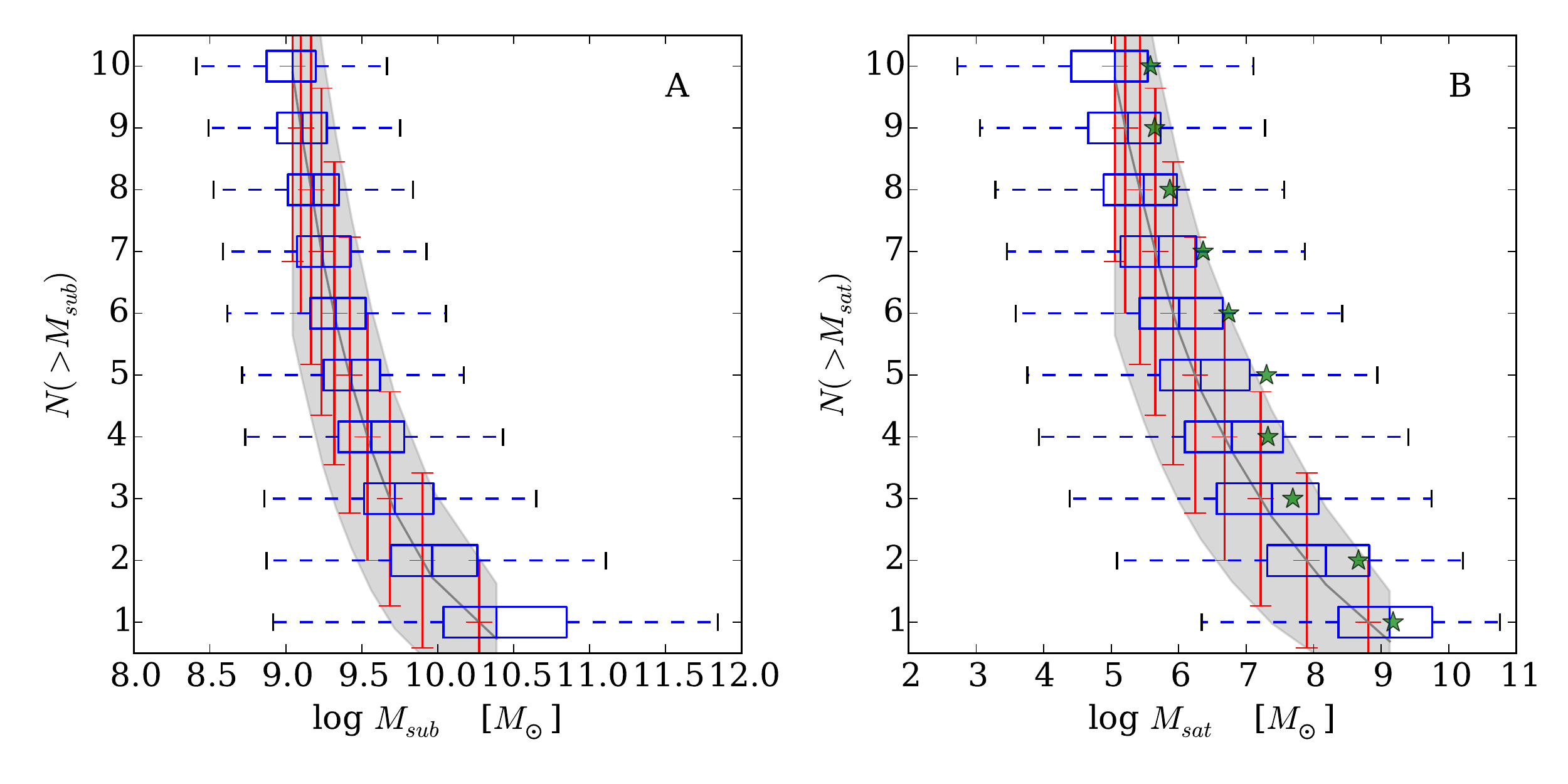}
\caption{The {\bf left} panel shows the subhalo mass function of the MW host halos 
in the \texttt{c125-2048} simulation.  
The {\bf right} panel shows the satellite stellar mass function predicted by the fiducial 
SAM using the merger trees of the MW host halos in the \texttt{c125-2048} simulation. 
In each panel, the horizontal whiskers show the distribution of the subhalo mass 
(or the stellar mass) of the $i^{\rm th}$ most massive subhalo (or satellite). 
The size of the box shows the quartiles of the distribution. 
The vertical bar in the middle of the box marks the median of the distribution in mass. 
The outer bars connected by dashed lines extend to the most extreme values. 
The red vertical error bars show the standard deviation in the mass functions assuming the distribution 
of the mass functions follows a Poisson distribution. 
The gray shaded region shows the 1$\sigma$ range of the simulation (SAM) predicted mass 
function. 
The dark gray line in the middle of the shaded region denotes the median of the distribution. 
In the satellite mass function panel ({\bf right}), the green stars denote the observational 
data of the stellar masses of the 10 most massive MW satellites from \citet{McConnachie2012a}.    
}
\label{fig:mfunc}
\end{center}
\end{figure*}

We study the probability distribution of the MW satellite mass function predicted by 
the $N$-body simulations and the SAM adopted in this paper. 
Two alternative models for the probability distribution function are tested here. 
One is the Poisson probability distribution, and the other is the Negative Binomial 
Distribution (NBD). 
\citet{Boylan-Kolchin2010a} found that the NBD provides a more accurate description 
for the distribution of the subhalo mass function predicted by $N$-body simulation. 
Similar findings have also been reported in independent studies 
\citep[e.g.,][]{Busha2011b, Cautun2014a}.  
To choose an accurate model for the distribution function of the satellite galaxies 
predicted by a galaxy formation model, we apply a fiducial SAM to a set of halo merger trees 
of MW-mass halos extracted from the cosmological simulation \texttt{c125-2048}.

We first extract the subhalo mass function from the simulation. 
Using a fiducial SAM, which was tuned to match the field galaxy stellar mass 
function of the local Universe \citep{Lu2014b}, we then make a prediction for the 
satellite galaxy stellar mass function. 
In Figure \ref{fig:mfunc}, we show the subhalo mass function on the left panel 
and the SAM predicted satellite stellar mass function on the right panel. 
Each blue horizontal box covers the lower to upper quartile values of the 
mass of the $i^{\rm th}$ subhalo (satellite), with a middle line indicating the 
median of the mass distribution. 
The whiskers extending from the boxes show the range of the masses in the sample. 
On the other hand, the gray shaded region covers the 1$\sigma$ range of the 
cumulative distribution function $N(>M)$ as a function of $\log M$. 
In comparison, we also show the expected 1$\sigma$ scatter of the mass functions
with red vertical error bars assuming the mass functions are Poisson distributed. 
We will show later that the real distribution of the mass functions quantitatively 
deviates from Poisson. 
Moreover, we also overplot the observed MW satellite stellar mass function in the right panel. 
Without further tuning the model against the data, the fiducial model prediction 
is in a good agreement with the data, with most of the data points encompassed by 
the $1\sigma$ region of the predicted distribution, leaving only 2 outliers with a small deviation. 
Although the fiducial model does not perfectly match the data, it is sufficiently close. 
We can use it as a representative model to explore the statistical behavior of predicted 
satellite populations. 

\begin{figure*}[htb!]
\begin{center}
\includegraphics[width=0.9\textwidth]{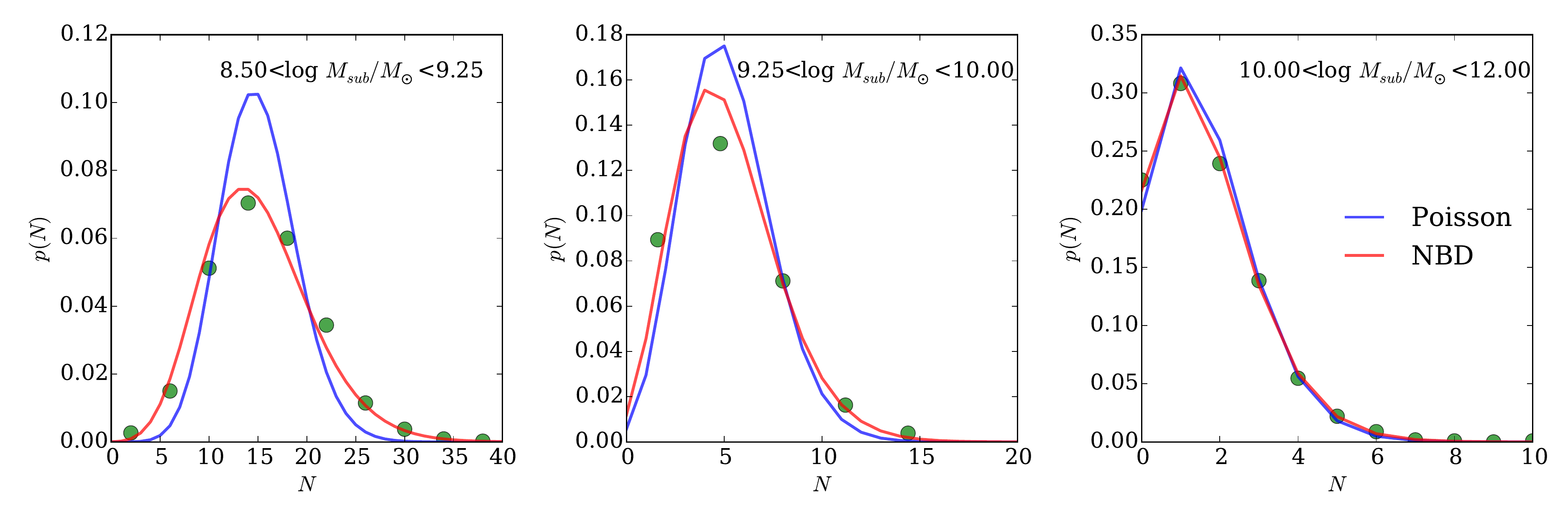}
\includegraphics[width=0.9\textwidth]{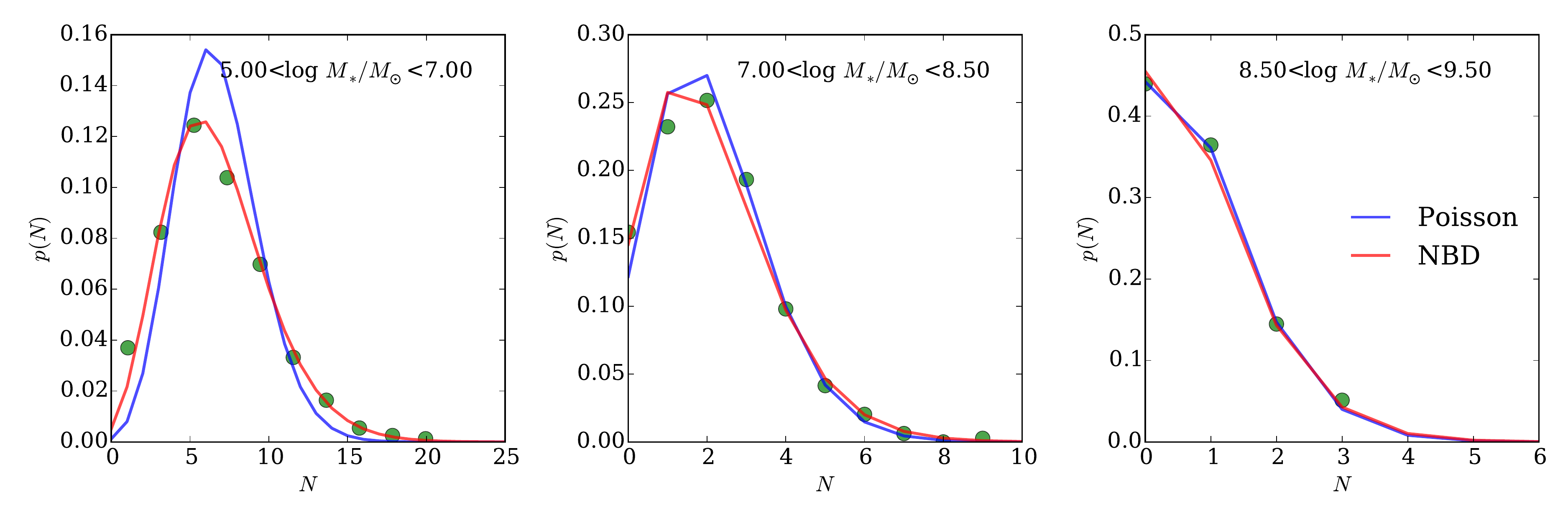}
\caption{{\bf Upper}: the distribution of the number of subhalos in three mass bins. 
{\bf Lower}: the distribution of the number of satellites in three stellar mass bins. 
In each panel, the green circles denote the distribution of subhalo mass function 
(satellite stellar mass function in the lower panels) 
predicted by the cosmological simulation and the fiducial SAM. 
The blue line and red lines show the Poisson distribution and the Negative Binomial 
distribution with the same mean as the simulation predicted, respectively. 
For the Negative Binomial distribution, we have assumed $s_{\rm I}=0.26$ in this figure.  
}
\label{fig:dist_smf}
\end{center}
\end{figure*}

The probability distribution function of the number count of subhalos is found 
to be better described by the Negative Binomial Distribution (NBD) \citep{Boylan-Kolchin2010a}, 
\begin{equation}
P(N | r, p) = \frac{ \Gamma(N+r)}{\Gamma(r) \Gamma(N+1)} p^r (1-p)^N\,,
\end{equation}
where $N$ is the number of subhalos per host in a given mass range, $\Gamma(x)\equiv(x-1)!$ is the Gamma function, 
and $r$ and $p$ are two parameters. 
This distribution function has also been used to describe the probability of the number 
of satellite galaxies in HOD models \citep[e.g.,][]{Berlind2002a}.
\citet{Boylan-Kolchin2010a} argued that the reason for this model to better describe the
distribution of the subhalo mass function than the Poisson distribution is because it 
captures the intrinsic non-Poisson scatter of the mass function.  
The authors defined a parameter $s_{\rm I}\equiv {{\sigma}_{\rm I} / \mu}$, 
the fractional scatter from the intrinsic scatter, $\sigma_{\rm I}$, 
with respect to the Poisson scatter, $\mu$. 
The two parameters in the NBD are then determined as 
\begin{equation}
p=\frac{1}{ 1 + s^2_{\rm I} \mu}\,, r= \frac{1}{s^2_{\rm I}}. 
\end{equation}

In Figure \ref{fig:dist_smf}, we show the distribution of subhalo number 
count of the MW-mass hosts in three different subhalo mass bins, which are noted in each panel. 
In agreement with previous studies, we also find that the NBD describes the subhalo number 
count distribution much better than the Poisson distribution. 
\citet{Boylan-Kolchin2010a} found that $s_{\rm I}\approx0.18$ yields a good fit 
to the subhalo mass function distribution predicted by the Millennium-II Simulation. 
We find, however, that a larger value for $s_{\rm I}$ is needed to fit the distribution predicted 
by our simulations. 
In the figure, we have used $s_{\rm I}=0.26$. 
When the expected number counts becomes smaller ($\bar{N}<4$), the NBD approaches the Poisson 
distribution, making it hard to distinguish between the two models. 
We also show the satellite galaxy number count distribution at three different stellar masses. 
As same as the subhalo mass function distribution, the number count distribution for given 
stellar mass can also be accurately described by the NBD. 
\citet{Busha2011b} found that adding an exponential tail to the NBD can better capture the 
distribution at very high $N$. 
Because the tail only covers a small fraction of the probability distribution, 
we ignore this part for keeping the model simple without losing accuracy above the level our 
inference can capture.  
Using the fiducial SAM, we predict stellar masses for the satellite galaxies and compute 
the stellar mass function for each MW host. 
We show the distribution of the number of satellite galaxies in three stellar mass bins in the lower 
panel of Figure \ref{fig:dist_smf}. 
Again, we use $s_{\rm I}=0.26$ for the NBD plotted in the figure to compare with the model predictions 
and find that the NBD matches the simulated distribution of the stellar mass function remarkably well.

\begin{figure*}[htb!]
\begin{center}
\includegraphics[width=0.9\textwidth]{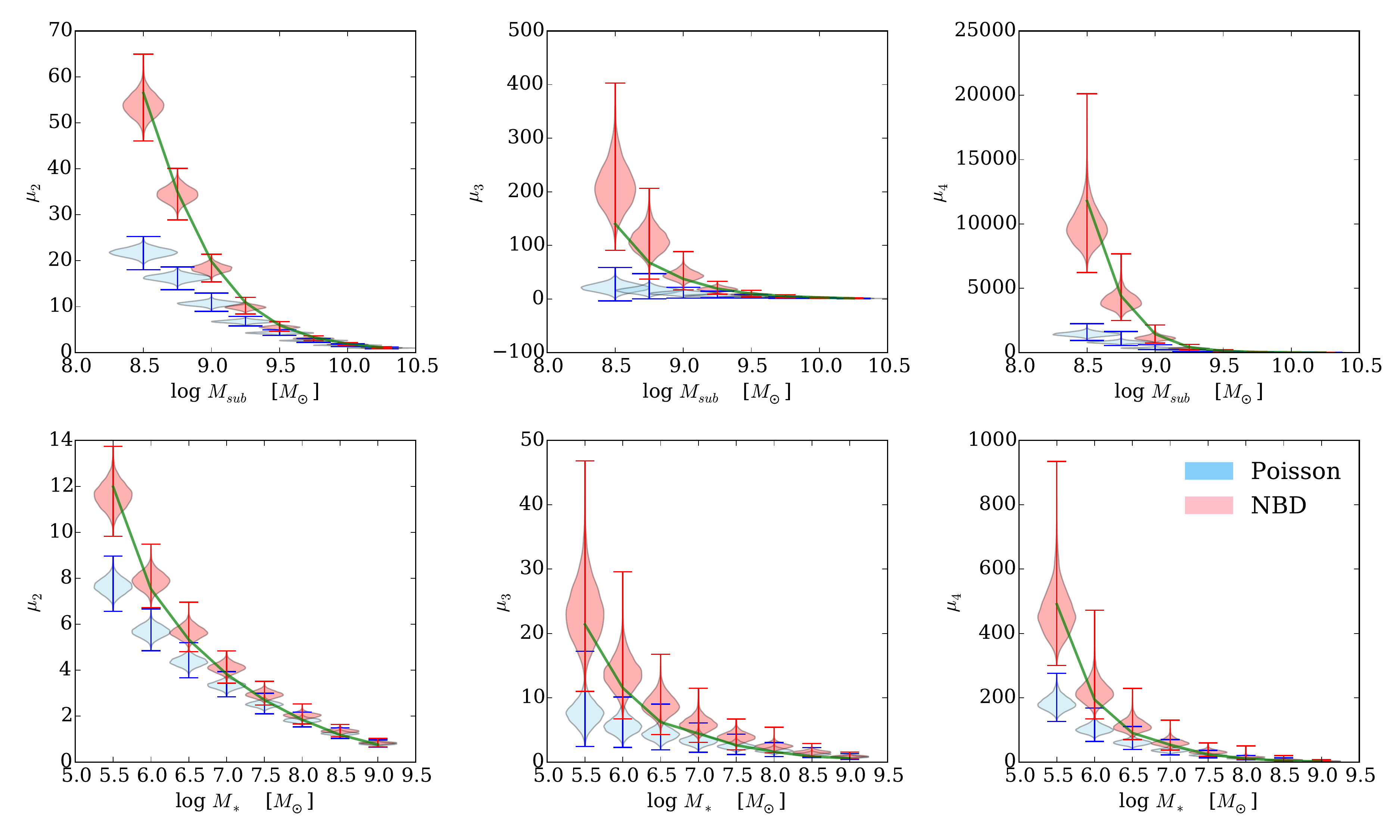}
\caption{The 2nd, 3rd, and 4th moments of the distribution of subhalo and satellite mass functions 
as a function of mass. 
The green line denotes the moments predicted by MW halos from the cosmological simulation.
The blue violin shapes denote the reference distributions of the moments as a function of mass 
assuming that the mass functions are Poisson distributed.
The red violin shapes denote the same reference distribution  assuming that the mass function 
follows the NBD. 
The simulation predicted mass functions have moments that are largely 
deviated from Poisson, especially for low masses, but consistent with the NBD. 
}
\label{fig:moments}
\end{center}
\end{figure*}

We now quantitatively test which model better matches the moments of 
the simulation predicted distribution of the subhalo and satellite mass functions. 
Using the large sample provided by the cosmological simulations, we compute 
the 2$^{\rm nd}$, 3$^{\rm rd}$, and 4$^{\rm th}$ order central moments of the subhalo mass function 
and the stellar mass function for given mass bins.
The $n^{\rm th}$ order central moment is defined as
\begin{equation}
\mu_n = E\left[\left(x - E[x]\right)^n\right]\,, 
\end{equation}
where function $E[x]$ is the expectation of $x$. 
1133 MW halos from the simulation are used.   
The simulation predicted moments for the subhalo mass function and 
the stellar mass function as functions of masses are shown as green lines in 
the upper panels and the lower panels of Figure \ref{fig:moments}, respectively. 
The moments monotonically increase with a decreasing mass. 
Based on the hypothesis that the distribution function is Poisson or NBD, 
we generate same number of Monte Carlo mass functions, assuming each mass 
function is a random realization of the assumed distribution function. 
We adopt the simulation predicted means to assign the expectation value for 
the Monte Carlo mass functions. 
Using the Monte Carlo mass functions, we can compute the same central moments 
for each given mass bin. 
We then replicate the Monte Carlo simulation 10000 times to obtain 10000 samples 
of the moments for each mass bin. 
We use these 10000 values to construct a reference distribution for each moment 
for any given mass bin.
In Figure \ref{fig:moments} we show the distribution of the Monte Carlo subhalo 
mass function and the satellite stellar mass function assuming the samples are Poisson 
distributed (blue) or follow the NBD (red). 
As one can see, the simulation predicted moments are largely deviated from the moments 
based on the Poisson distribution, especially in the low-mass bins. 
These Monte Carlo simulations suggest that while the simulation predicted distribution 
is still consistent with Poisson at high-mass bins, 
it is clearly inconsistent with Poisson for low mass bins. 
In contrast, the NBD moments matches the moments of the simulated mass functions remarkably well.


\end{document}